\documentclass{elsart}
\usepackage{graphicx}
\usepackage {subfigure,epsfig}

\usepackage {amsmath} \usepackage{amssymb} \usepackage{cite} %\usepackage{amsthm}
\numberwithin{equation}{section} %\numberwithin{table}{section}
\begin{document}
\begin{frontmatter}

\title{Power and non-power expansions of the solutions for the fourth-order analogue to the second Painlev\'{e} equation}
\author{Maria V. Demina,}
\author{Nikolai A. Kudryashov\corauthref{cor}}
\corauth[cor]{Corresponding author.}
\address{Department of Applied Mathematics, Moscow  Engineering
 and Physics Institute (State university), 31 Kashirskoe Shosse,
115409 Moscow, Russian Federation} \ead{kudryashov@mephi.ru}

\begin{abstract}
Fourth - order analogue to the second Painlev\'{e} equation is
studied. This equation has its origin in the modified Korteveg - de
Vries equation of the fifth order when we look for its self -
similar solution. All power and non - power expansions of the solutions for the fouth
- order analogue to the second Painlev\'{e} equation near points
$z=0$ and $z=\infty$ are found by means of the power geometry
method. The exponential additions to solutions of the equation
studied are determined. Comparison of the expansions found with
those of the six Painlev\'{e} equations confirm the conjecture that the
fourth - order analogue to the second Painlev\'{e} equation defines
new transcendental functions.

\end{abstract}

\end{frontmatter}

\section{Introduction.}
More than one century ago Painlev\'{e} and his collaborators
analyzed a certain class of the second-order nonlinear ordinary
differential equations (ODE). In fact Painlev\'{e} examined two
problems. The first one was to find all the second-order canonical
equations with the solutions without movable critical points. The
second problem was to pick out those equations which have solutions
defining new special functions. The former was posed by Picard in
1876. While the latter was mentioned by A. Pouancare and L. Fuchs in
1884.

As a result of investigations Painlev\'{e} and his school discovered
six second-order nonlinear ODEs which solutions could not be
expressed in terms of known elementary or special functions.
Nowadays they go under the name of the Painlev\'{e} equations and
their solutions are called the Painlev\'{e} transcendents.

For a long period of time the Painlev\'{e} equations were regarded
as nothing more than a curious moment in the theory of differential
equations and were beyond the center of  scientific circles
attention. However since the sixtieth of the twentieth century a
number of works appeared where it was shown that the Painlev\'{e}
transcendents arose in the models describing physical phenomena as
frequently as many other special functions \cite{Ablowitz01,
Barouch03, Brezin04, De_Boer05, Ablowitz06, Hall07, Chandrasecar08,
Kudryashov01, Kudryashov02}. This fact caused a significant interest
to the studying of their properties and set a problem to find other
nonlinear ODEs defining new transcendents.

It is necessary to mention that exact solutions of many partial
nonlinear differential equations (such as the Korteweg-de Vries
equation, the modified Korteweg-de Vries equation, the nonlinear
Schr\"{o}dinger equation, the sine-Gordon equation and so on)
solvable by the inverse scattering transform can be expressed via
the Painlev\'{e} transcendents and it is one of their most important
applications \cite{Ablowitz01, Ablowitz06, Kudryashov01,
Kudryashov03, Hone01}. In this connection it is quiet natural to
suppose that the partial solutions of exactly solvable equations of
the order higher than the order of previously mentioned equations
also can be expressed in terms of the transcendents defined as the
solutions of nonlinear ODEs. This idea was developed in the work
\cite{Kudryashov03}, where an hierarchy of the first Painlev\'{e}
equation was introduced. Nowadays the higher analogues of the
Painlev\'{e} equations are intensively studied \cite{Kudryashov03,
Kudryashov04,Kudryashov05, Kudryashov06, Kudryashov07, Kudryashov08,
Kudryashov09, Kudryashov10, Kudryashov11, Kudryashov12,
Kudryashov13, Kudryashov14, Kudryashov15, Kudryashov16,
Airault01,Clarkson01, Clarkson02, Cosgrove01, Joshi01, Hone01,
Hone02, Gordoa01, Gordoa02, Flashka01, Kawai01, Mugan01, Mugan02,
Mugan03, Mugan04, Nijhoff01, Pickering01, Shimomura01}.

Taking after \cite{Kudryashov03, Airault01, Hone01, Flashka01} we
will show that a solution of the modified Korteveg - de Vries
equation of the fifth order
\begin{equation}
\label{1.1}u_t+u_{xxxxx}-10u^{2}u_{xxx}-40u_{x}u_{xx}-10u_{x}^{3}+30u^{4}u_{x}=0
\end{equation}
is defined by the solution of the fourth-order analogue of the
second Painlev\'{e} equation. Let us look for the solution of the
equation \eqref{1.1} in the form
\begin{equation}
\label{1.2}u(x,t)=\frac{1}{(5t)^{1/5}} \,w(z),\,\,\,\,\,z=\frac
x{(5t)^{1/5}}.
\end{equation}
After substitution of \eqref{1.2} into \eqref{1.1} and integrating
once by $z$ we get the equation
\begin{equation}
\label{1.3}f(z,w)\stackrel{def}{=}w_{zzzz} - 10w^2w_{zz} - 10ww^2_z
+ 6 w^5  - zw - \alpha =0
\end{equation}
This equation has a number of properties similar to that of
Painlev\'{e} equations. More exactly it possesses B\"{a}cklund
transformations \cite{Kudryashov05, Clarkson01, Hone01}, a Lax pair
\cite{Kudryashov05}, rational and special solutions at certain
values of the parameter $\alpha$ \cite{Kudryashov05, Kudryashov11,
Clarkson01, Hone01}. The Caushy problems for the equation
\eqref{1.3} can be solved by the isomonodromic transfom method. It
is likely that the equation \eqref{1.3} defines new transcendental
functions as Painlev\'{e} equations do. However the exact proof of
this statement yet is an open question. In this connection an
important step of the investigation of the equation \eqref{1.3}
properties is the finding of all its solution asymptotics.

The aim of this work is to calculate all power and non-power
asymptotics, power and exponential expansions for the solutions of
the equation \eqref{1.3} with a help of the power geometry method
\cite{Bruno01, Bruno02, Bruno03}.

The problem definition is

1) to find all power asymptotics of its solutions. If a solution of
the equation supposing that $z\rightarrow 0$ or $z\rightarrow
\infty$ can be presented in the form
\begin{equation}
\label{1.4} w(z)= c_r\,z^{r}+ o(|z|^{r+\varepsilon}),
\end{equation}
where the coefficient $c_r=const\in \mathbb{C}$, $c_r \neq 0$, the
exponents $r,\varepsilon \in \mathbb{R}$, $\varepsilon \omega < 0$
and
\begin{equation*}
\omega=\left\{
\begin{gathered}
-1,\,z\rightarrow 0;\\
\hfill 1,z\rightarrow\infty,
\end{gathered}
\right.
\end{equation*}
then the power asymptotic of the solution \eqref{1.4} is
\begin{equation}
\label{1.5} w(z)= c_r\,z^{r}.
\end{equation}

2) to calculate all power expansions of its solutions given by
\begin{equation}
\label{1.6} w(z)= c_r\,z^{r}+ \sum_{s}\,c_s\,z^{s},
\end{equation}
where $s>r$ if $z\rightarrow\,0$ and $s<r$ if
$z\rightarrow\,\infty$.

3) to find for the studied equation all non-power asymptotics, i.e.
functions $u(z)$ that are connected with the solutions of the
equation \eqref{1.3} in the way
\begin{equation}
\label{1.7} w(z)= u(z)\left[1+o(|z|^{\omega \varepsilon}
|u(z)|^{\omega_1\varepsilon})\right]
\end{equation}
for some $\varepsilon<0$ where
\begin{equation*}
\omega_1=\left\{
\begin{gathered}
\,\,0,\,u(z)\rightarrow const\neq0;\\
-1,\hfill u(z)\rightarrow0;\\
\,\,1,\hfill u(z)\rightarrow\infty.
\end{gathered}
\right.
\end{equation*}

4) to find all exponential additions to power expansions of the
solutions which correspond to the exponentially close solutions. In
other words it is required to find functions
\begin{equation}
\label{1.8} v(z)= \exp \left[\sum \gamma_sz^s\right],\,\,\,\omega
s<const,
\end{equation}
which, added to the solution $w(z)+v(z)$, correspond to the power
expansions.

This paper outline is as follows. The general properties of the
fourth - order analogue to the second Painlev\'{e} equation are
discussed in section 2. Power expansions corresponding to the apexes
and to the edges are given in sections 3, 4, 5, 6 and 7. In section
8 we consider the general properties of the equation studied in the
special case (at $\alpha=0$). Corresponding power and non-power
expansions of solutions are presented in sections 9, 10 and 11. The
three-level exponential additions are determined in sections 12, 13,
14, 15,16, 17 and 18. The results of the work are gathered in
section 19.

\section{General properties of the equation \eqref{1.3} at $\alpha \neq 0$.}
The following points correspond to the monomials of the studied
equation: $M_1=(-4,1),\,\,\, M_2=(-2,3),\,\,\,M_3=(-2,3),\,\,\,
M_4=(0,5),\,\,\, M_5=(1,1),\,\,\, M_6=(0,0)$. The carrier $S(f)$ of
the equation contains five points $Q_1=M_1,\,\,\, Q_2=M_4,\,\,\,
Q_3=M_5 ,\,\,\, Q_4=M_6$ è $Q_5=M_2=M_3$. Their convex hull is the
quadrangle with four apexes $\Gamma_j^{(0)}=Q_j\,(j=1,2,3,4)$ and
four edges
$\Gamma_1^{(1)}=[Q_1,Q_2],\,\Gamma_2^{(1)}=[Q_2,Q_3],\,\Gamma_3^{(1)}=[Q_3,Q_4],\,\Gamma_4^{(1)}=[Q_1,Q_4]$
(fig. 1).
\begin{figure}[h]
 \centerline{\epsfig{file=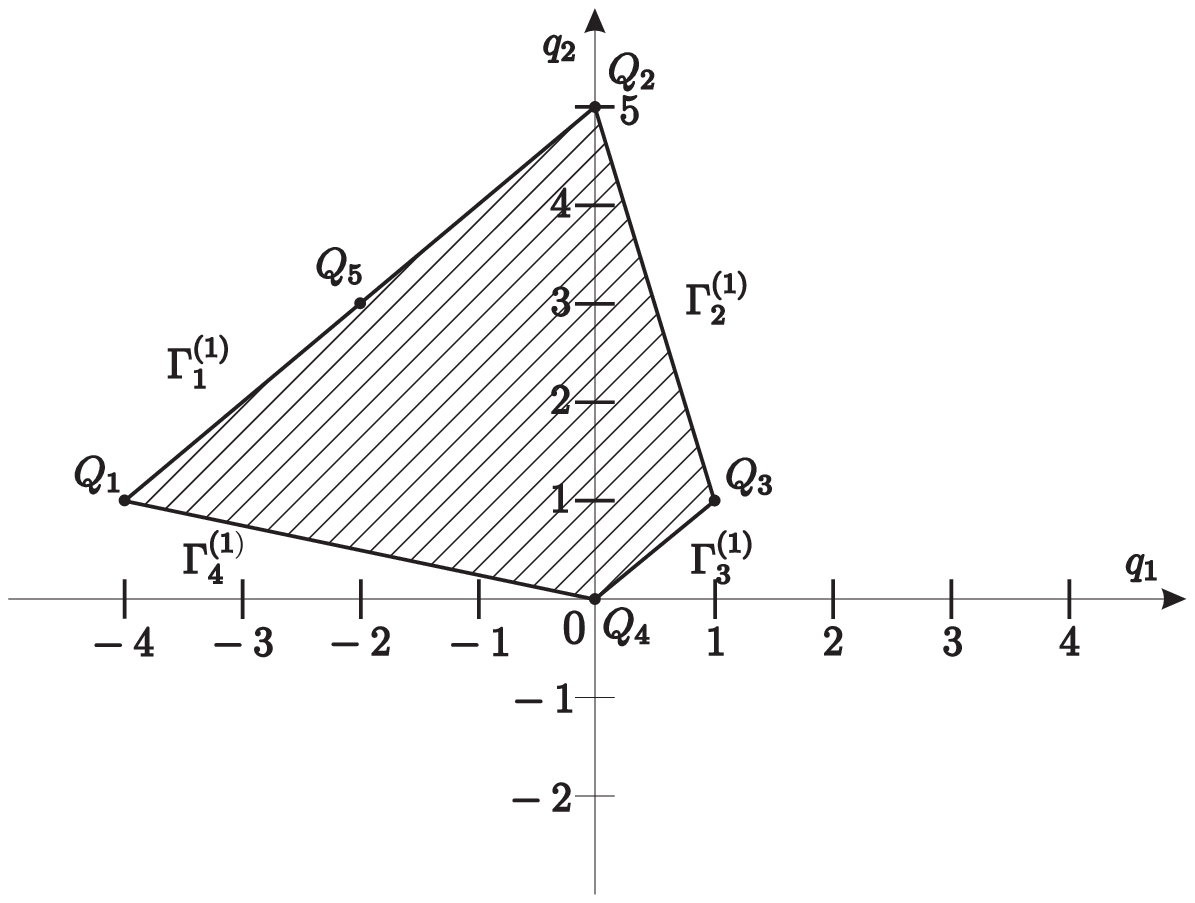,width=120mm}}
 \caption{}\label{fig:z_post}
\end{figure}

The external normal vectors $N_j\,(j=1,2,3,4)$ to edges
$\Gamma_j^{(1)}\,(j=1,2,3,4)$ are $N_1=(-1,1),\, N_2=(4,1),\,
N_3=(1,-1),\, N_4=(-1,-4)$. They form the normal cones $U_j^{(1)}$
of edges $\Gamma_j^{(1)}$
\begin{equation}
\label{2.1}U_j^{(1)} =\{P=\lambda N_j,\,\,\, \lambda>0\},\,\,\,
j=1,2,3,4.
\end{equation}
The normal cones $U_j^{(0)}$ of the apexes $\Gamma_j^{(0)}=Q_j\,\,
(j=1,\,2,\,3,\,4)$ are the angles between the edges that adjoin to
the apex (fig. 2).
\begin{figure}[h]
 \centerline{\epsfig{file=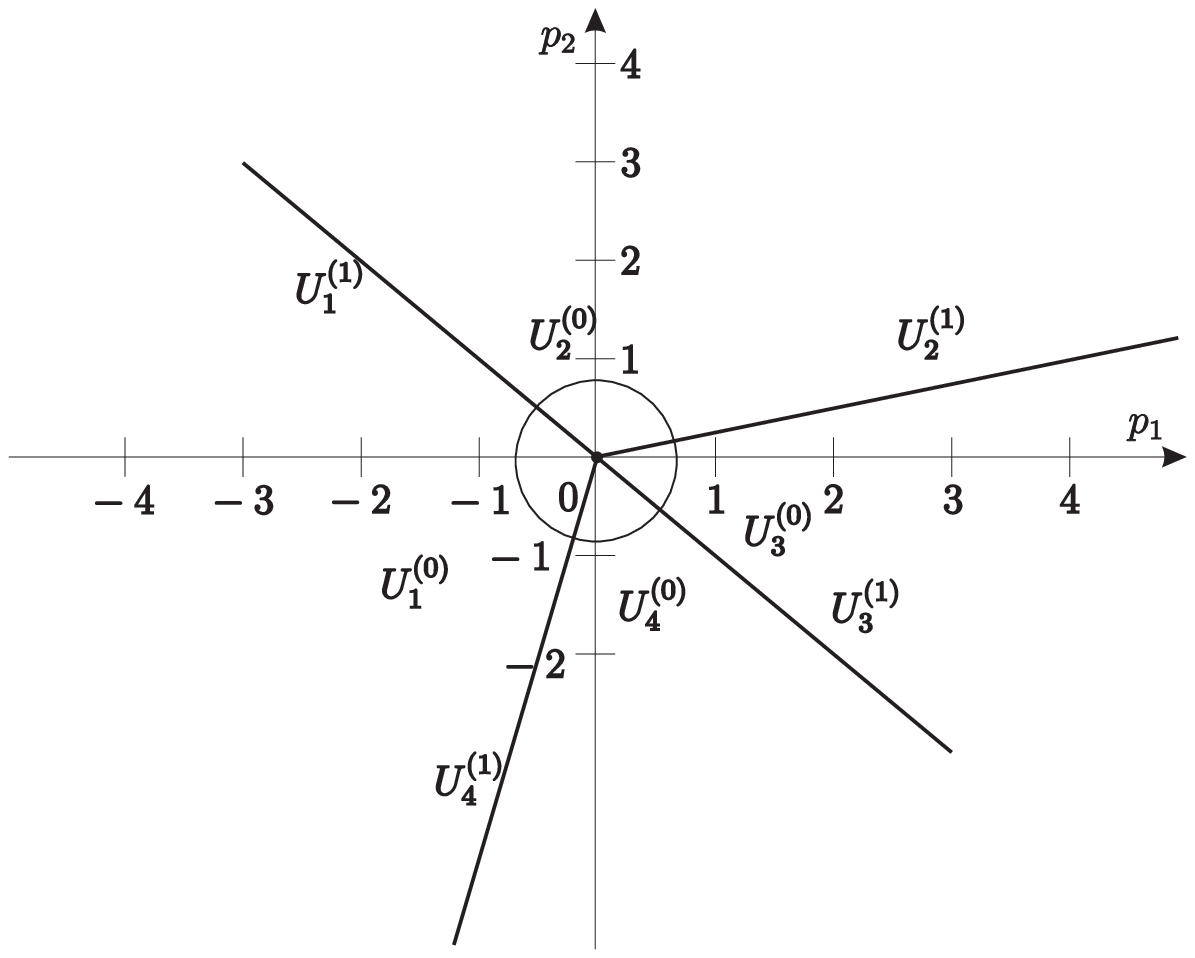,width=120mm}}
 \caption{}\label{fig:z_post}
\end{figure}

The carrier of the equation \eqref{1.1} lies in the lattice
$\mathbb{Z}$ with the basis $B_1=Q_1=(-4,1),\,B_2=Q_3=(1,1)$.
Expressing the points $Q_2,\,Q_5$ through the measuring vectors we
get $Q_2=4B_2+B_1$, $Q_5=B_1+2B_2$. Examining the reduced equations
that correspond to the bounds $\Gamma_j^{(d)}\,\, (d=0,1;\,\,
j=1,\,2,\,3,\,4)$ we will study the solutions of the equation
\eqref{1.1}. Note that the reduced equations which correspond to the
apexes $\Gamma_2^{(0)},\,\Gamma_3^{(0)},\,\Gamma_4^{(0)}$
\begin{equation}
\begin{gathered}
\label{2.4}\hat{f}_2^{(0)}\stackrel{def}{=}6w^5=0,\\
\hat{f}_3^{(0)}\stackrel{def}{=} -zw=0,\\
\hat{f}_4^{(0)}\stackrel{def}{=}-\alpha=0
\end{gathered}
\end{equation}
are the algebraic ones and that is why they do not have non-trivial
power or non-power solutions.

\section{Power expansions, corresponding to the apex $\Gamma_1^{(0)}$ at $\alpha \neq 0$.}
The apex $\Gamma_1^{(0)}$ defines the following reduced equation
\begin{equation}
\label{3.1}\hat{f}_1^{(0)}\stackrel{def}{=} w_{zzzz}=0.
\end{equation}
Let us find the reduced solutions
\begin{equation}
\label{3.2}w=c_r z^r,\,\,\, c_r\neq0
\end{equation}
for $\omega (1,r) \in U_1^{(0)}$. Since $p_1 < 0$ in the cone
$U_1^{(0)}$ then $\omega=-1, \,\,\, z\rightarrow 0$ and the
expansions will be the ascending power series of $z$. Substituting
\eqref{3.2} into $\hat{f}_1^{(0)}$ and canceling the result by
$z^{r-4}$ we get the characteristic equation
\begin{equation}
\label{3.3}\chi(r)\stackrel{def}{=} r(r-1)(r-2)(r-3)=0
\end{equation}
with the four roots $r_1=0,\,r_2=1,\,r_3=2,\,$ and $r_4=3$. Further
we shall examine them separately.

Vector $R=(1,0)$ that corresponds to the root $r_1=0$ being
multiplied by $\omega$ belongs to the cone $U_1^{(0)}$. Thus we
obtain the family $F_1^{(0)}\,1$ of power asymptotics $w=c_0$ with
the arbitrary constant $c_0 \neq 0$. The first variation of the
equation \eqref{3.1}
\begin{equation}
\label{3.4}\frac{\delta\hat{f}_1^{(0)}}{\delta w} = \frac{d^4}{dz^4}
\end{equation}
gives the operator
\begin{equation}
\label{3.5}L(z)=\frac{d^4}{dz^4} \neq 0
\end{equation}
with the characteristic polynomial
\begin{equation}
\label{3.6}\nu(k) =z^{4-k} L(z) z^k =k(k-1)(k-2)(k-3).
\end{equation}
The equation $ \nu(k)=0 $ has four roots $k_1=0,\,\,\, k_2=1,\,\,\,
k_3=2,\,\,\,$è $k_4=3$. As $\omega=-1$ and $r=0$ then the cone of
the problem is $\mathcal{K}=\{k>0\}$. It contains
$k_2=1,\,k_3=2,\,k_3=3$ which are the critical numbers. Let us
investigate the equation obtained from the equation \eqref{1.3}
after changing of variables $w(z)=c_{r_1}z^{r_1}+y(z)$
\begin{equation}
\label{3.7}g(z,y) \stackrel{def}{=} f(z,c_{r_1}z^{r_1}+y).
\end{equation}
Its carrier $ S(g)$ contains the points belonging to $ S(f)$ and
seven other points: $ (-2,2),\, (-2,1),\, (0,4),\, (0,3),\, (0,2),\,
(0,1),\, (1,0)$. Displacing all these points on the vector $(-4,1)$
we get $(2,1)$, $ (2,0)$, $(4,3)$, $(4,2)$, $(4,1)$, $(4,0)$,
$(5,1)$, $(4,4)$, $(0,0)$, $(4,-1)$, $(5,0)$. The set $ S^{'}_{+}$
of the finite sums of the vectors based on these points intersects
with the line $ q_{2}=-1 $ at the points
\begin{equation}
\label{3.8}\mathbb{K}=\{k=4+n,\,\,\,n\in N\cup{\{0\}}\}.
\end{equation}
The critical numbers do not lie in this set that is why the indexes
$s$ in the expression \eqref{1.6} belong to the set
$\mathbb{K}(1,2,3)$.
\begin{equation}
\label{3.9}\mathbb{K}(1,2,3)=\{k=n,\,\,\,n\in N\ \}.
\end{equation}
Thus the expansion of the solution corresponding to the reduced
equation \eqref{3.1} at $r=0$ is the following
\begin{equation}
\label{3.10}w(z)=c_0+\sum_{k\in\mathbb{K}(1,2,3)}c_kz^k\equiv c_0 +
c_1 z + c_2z^2 + c_3z^3 + \sum^{\infty}_{k=4} c_k z^k
\end{equation}
where all the coefficients are constants, $c_0\neq
0,\,c_1,\,c_2,\,c_3$ are arbitrary ones and $c_k \,\, (k\geq 4)$ are
uniquely defined. Denote this family as $G_1^{(0)}\,1$. Taking into
account eight members, the expansion \eqref{3.10} can be presented
in the form
\begin{equation*}\begin{gathered}
w(z)=c_{{0}}+c_{{1}}\,z+c_{{2}}\,{z}^{2}+c_{{3}}\,{z}^{3}+\left(
\frac1{24} \,\alpha-\frac14\,{c_{{0}}}^{5}+{\frac
{5}{12}}\,c_{{0}}{c_{{1}}}^{2}+\frac56 \,{c_{{0}}}^{2}c_{{2}}
\right) {z}^{4}+
\\+\left(\frac1{12}\,{c_{{1}}}^{3}+\frac12\,{c_{{0}}}^{2}c_{{3}}-\frac14\,{c_{{0}}}^{4}
c_{{1}}+{\frac{1}{ 120}}\,c_{{0}}+\frac23\,c_{{0}}c_{{1}}c_{{2}}
\right) {z}^{5}+
\\+\left( \frac29\,c_{{0}}{c_{{2}}}^{2}+\frac13
\,{c_{{0}}}^{2}c_{{4}}+\frac12\,c_{{0}}c_{{1}}c_{{3}}+{\frac
{1}{360}}\,c_ {{1}}+{\frac
{7}{36}}\,c_{{2}}{c_{{1}}}^{2}-\frac1{12}\,{c_{{0}}}^{4}c_{{2}}
-\frac16\,{c_{{0}}}^{3}{c_{{1}}}^{2} \right) {z}^{6}+
\\+\left( {\frac
{5}{21}}\,{c_{{0}}}^{2}c_{{5}}+{\frac {1}{840}}\,c_{{2}
}-\frac1{28}\,{c_{{0}}}^{4}c_{{3}}+{\frac{8}{21}}\,c_{{0}}c_{{1}}c_{{4}}+\frac13\,c_{{0}}c_{{2}}c_{{3}}+
\right. \\
\left. + \frac17\,{c_{{2}}}^{2}c_{{1}}-\frac17\,{c_{{0}}}^{3}c
_{{1}}c_{{2}} +{\frac
{13}{84}}\,{c_{{1}}}^{2}c_{{3}}-\frac1{14}\,{c_{{0}}}^{
2}{c_{{1}}}^{3} \right){z}^{7}+...
\end{gathered}\end{equation*}

The cone of the problem is $\mathcal{K}=\{k>1\}$ in the case
$r_2=1$. Consequently there are two critical numbers $k_2=2,\,
k_3=3$. Likewise the previous case we find the solution expansion
\begin{equation}
\label{3.11}w(z)=c_1z + c_2z^2 + c_3z^3 + \sum^{\infty}_{k=4} c_kz^k
\end{equation}
that is generated by the power asymptotic $\mathcal{F}_1^{(0)}2:
\,\,\, w= c_1\,z$. Here $c_1\neq 0,\, c_2$ and $c_3$ are arbitrary
constants.

For the root $r_2=2$ the cone of the problem is
$\mathcal{K}=\{k>2\}$. It contains $k_3=3$ that is the unique
critical number. The power expansion corresponding to the asymptotic
$\mathcal{F}_1^{(0)}3: \,\,\, w= c_2\,z^{2}$ takes the form
\begin{equation}
\label{3.12}w(z)=c_2z^2 +c_3z^3 + \sum^{\infty}_{k=4} c_kz^k.
\end{equation}
Again $c_2\neq 0,\,c_3$ are arbitrary constants. Denote this family
as $\emph{G}_1^{(0)}3$.

For the root $r_3=3$ the cone of the problem is
$\mathcal{K}=\{k>3\}$. There are no critical numbers in this case.
The expansion of the solution $\emph{G}_1^{(0)}4$ corresponding to
the power asymptotic $\mathcal{F}_1^{(1)}4: \,\,\, w= c_3\,z^{3}$
can be written as
\begin{equation}
\label{3.13}w(z)=c_3z^3 + \sum^{\infty}_{k=4} ñ_kz^k.
\end{equation}

Obtained expansions converge for small $|z|$. The existence and the
analyticity of the expansions \eqref{3.10}, \eqref{3.11},
\eqref{3.12}, \eqref{3.13} follow from the Cauchy theorem. The apex
does not define non-power asymptotics and exponential additions.

\section{Power expansions, corresponding to the edge $\Gamma_1^{(1)}$ at $\alpha \neq 0$.}

The edge $\Gamma_1^{(1)}$ is characterized by the reduced equation
\begin{equation}
\label{4.1}\hat{f}^{(1)}_1 (z,w) \stackrel{def}{=}w_{zzzz}
-10\,w^2\,w_{zz} -10\,w\,w_{z}^2+6\,w^5=0
\end{equation}
and the normal cone $U^{(1)}_1=\{-\lambda(1,-1),\,\lambda>0\}$.
Therefore $\omega=-1$, i.e. $z\rightarrow 0$ and $r=-1$.
Consequently the solution of the equation \eqref{4.1} should be
looked for in the form $w=c_{-1} z^{-1}$. From the determining
equation
\begin{equation}
\label{4.2}c_{-1}(3\,c_{-1}^4 -15\,c_{-1}^2 +12) =0
\end{equation}
we find the values of coefficient $c_{-1}$ (not equal to zero):
$c_{-1}^{(1)} =1,\, c_{-1}^{(2)}=-1,\, c_{-1}^{(3)}=2,\,
c_{-1}^{(4)}=-2$. Hence we have four families of power asymptotics
\begin{equation}
\label{4.3}\mathcal{F}_1^{(1)}1: \,\,\,
w=z^{-1};\,\,\quad\,\mathcal{F}_1^{(1)}2: \,\,\, w=-z^{-1};
\end{equation}
\begin{equation}
\label{4.4}\mathcal{F}_1^{(1)}3: \,\,\,
w=2z^{-1};\,\,\quad\,\mathcal{F}_1^{(1)}4: \,\,\, w=-2z^{-1}.
\end{equation}
Let us compute the corresponding critical numbers. The first
variation
\begin{equation}
\label{4.5}\frac{\delta f_2^{(1)}}{\delta w}=\frac{d^4}{dz^4}
-20w_{zz}w -10w^2\frac{d^2}{dz^2} -10w_{z}^2 -10\,ww_z \frac d{dz} +
30\,w^4
\end{equation}
applied to the the solutions \eqref{4.4} yields the operator
\begin{equation}
\label{4.6}\mathcal{L}^{(1,2)}(z)
=\frac{d^4}{dz^4}-\frac{10}{z^2}\frac{d^2}{dz^2}+\frac{20}{z^3}
\frac{d}{dz} - \frac{20}{z^4}.
\end{equation}
Its characteristic polynomial is
\begin{equation}
\label{4.7}\nu(k ) ={k}^{4}-6\,{k}^{3}+{k}^{2}+24\,k-20.
\end{equation}
The equation $\nu(k)=0$ has the roots $k_1=-2,\,k_2=1,\,
k_3=2,\,k_4=5$.

With reference to the solutions \eqref{4.5} variation \eqref{4.6}
gives the operator
\begin{equation}
\label{4.8}\mathcal{L}^{(3,4)}(z) =\frac{d^4}{dz^4} - \frac{40}{z^2}
\frac{d^2}{dz^2} +\frac{80}{z^3} \frac{d}{dz} - \frac{280}{z^4},
\end{equation}
Its characteristic polynomial
\begin{equation}
\label{4.9}\nu(k)={k}^{4}-6\,{k}^{3}-29\,{k}^{2}+114\,k+280,
\end{equation}
has the roots $k_1=-4,\, k_2=-2,\, k_3=5,\, k_4=7$. The cone of the
problem looks like
\begin{equation}
\label{4.10} \mathcal{K}=\{k>-2\}.
\end{equation}
Thus for the power asymptotics \eqref{4.4} there are three critical
numbers (three roots of the characteristic polynomial belong to the
cone of the problem) and for the power asymptotics \eqref{4.5} there
are only two critical numbers. The shifted carrier of the power
asymptotics \eqref{4.4}, \eqref{4.5} is the vector $(-1,-1)=-B_{2}$.
It belongs to the lattice that consists of the points $Q=
m(-4,1)+l(1,1)$ where m and l  are the whole numbers. These points
intersect with the line $q_{2}=-1$ if $m+l=-1$, i.e. $l=-m-1$. As
the cone of the problem is \eqref{4.11} then
\begin{equation}
\label{4.11}\textbf{K}=\{k=-1+5m,\,\,\,m\in \mathbb{N}\}.
\end{equation}
Now we will look for the solution expansions generated by the
families \eqref{4.4}. The sets $\textbf{K}(1)$, $\textbf{K}(1,2)$
and $\textbf{K}(1,2,5)$ can be written as
\begin{equation} \begin{gathered}
\label{4.12}\textbf{K}(1)=\{k=-1+5m+2l,\,\,l,m\in
\mathbb{N}\cup\{0\},\,\,l+m\neq\,0\}=\\
=\{1,3,4,5,6,7,8,...\};
\end{gathered}\end{equation}
\begin{equation}\begin{gathered}
\label{4.13}\textbf{K}(1,2)=\{k=-1+5m+2l+3k,\,l,m,k\in
\mathbb{N}\cup\{0\},\,\,m+l+k\neq\,0\}=\\
=\{1,2,3,4,5,6,7,8,...\};
\end{gathered} \end{equation}
\begin{equation}\begin{gathered}
\label{4.14}\textbf{K}(1,2,5)=\{k=-1+5m+2l+3k+6n,\,n,m,k,l \in
\mathbb{N}\cup\{0\} ,\,\,m+n+k+l\neq\,0\}=\\
=\{n  ,\,\,n \in \mathbb{N}\}.
\end{gathered}\end{equation}
In this case the solution expansions are
\begin{equation}\begin{gathered}
\label{4.15}w(z)=\frac{c^{(1,2)}_{-1}}{z}+
\sum^{}_{k\,\in\,\textbf{K}(1,2,5)} c_{k}^{(1,2)}\,z^{k}.
\end{gathered}\end{equation}
Denote these families as $G_{1}^{(1)}1$ and $G_{1}^{(1)}2$.
Evidently the critical number $1$ does not belong to the set
$\textbf{K}$ that is why the compatibility condition for $c_1$ holds
automatically and in consequence $c_1$ is an arbitrary constant. The
critical number $2$ also does not belong to the sets $\textbf{K}$,
$\textbf{K}(1)$ and thus $c_2$ is an arbitrary constant. However the
critical number $5$ lies in the sets $\textbf{K}(1)$ and
$\textbf{K}(1,2)$. As a result it is necessary to verify that the
compatibility condition for $c_5$ is true. It turns out that it is
so then $c_5$ is an arbitrary constant. The three-parametric power
expansion that corresponds to  the power asymptotics \eqref{4.4} (at
$c_{-1}=1$) is the following
\begin{equation*}\begin{gathered}
\label{}
w(z)=\frac{1}{z}+c_{{1}}z+c_{{2}}{z}^{2}+\frac52\,{c_{{1}}}^{2}{z}^{3}+
\left( \frac53 \,c_{{1}}c_{{2}}-\frac1{36}\,\alpha-\frac1{36}
\right) {z}^{4}+c_{{5}}{z}^{5}+
\\ +\left( \frac58\,c_{{2}}c_{{3}}+{\frac
{37}{48}}\,{c_{{1}}}^{2}c_{{2}}-{ \frac
{7}{288}}\,c_{{1}}\alpha-{\frac {13}{720}}\,c_{{1}} \right) {z}^
{6}+
\\+ \left( {\frac
{ 17}{27}}\,c_{{1}}{c_{{2}}}^{2}-{\frac
{17}{1620}}\,c_{{2}}-\frac1{18}\,{c_{{1}}}^{4}-{\frac
{1}{81}}\,c_{{2}}\alpha-{\frac {
5}{54}}\,{c_{{1}}}^{2}c_{{3}}+\frac59\,c_{{1}}c_{{5}}+{\frac
{11}{54}}\,{c _{{3}}}^{2} \right) {z}^{7}+...
\end{gathered}
\end{equation*}
Here $c_i\equiv c_i^{(1)},\,i=1,2,5$.

The carrier of the power expansions generated by the families
\eqref{4.5} is defined by the sets
\begin{equation}\begin{gathered}
\label{4.16}\textbf{K}(5)=\{k=-1+5m+6l,\,l,m\in
\mathbb{N}\cup\{0\},\,\,m+l\neq\,0\}=\\
=\{4,5,9,10,11,14,15,16,17,19,...\};
\end{gathered}\end{equation}
\begin{equation}\begin{gathered}
\label{4.17}\textbf{K}(5,7)=\{k=-1+5m+6l+8k,\,m,l,k\in
\mathbb{N}\cup\{0\},\,\,m+l+k\neq\,0\}=\\
=\{4,5,7,9,10,11,12,13,14,...\}.
\end{gathered}\end{equation}
The solution expansions in this case are
\begin{equation}\begin{gathered}
\label{4.18}w(z)=\frac{c^{(3,4)}_{-1}}{z}+ \sum^{}_{k\,\, \in \,\,
\textbf{K}(5,7)} c_{k}^{(3,4)}\,z^{k}.
\end{gathered}\end{equation}
Denote these families as $G_{1}^{(1)}3$, $G_{1}^{(1)}4$. Obviously the
critical numbers $5$ and $7$ do not lie in the set $\textbf{K}$ and
besides the critical number $7$ does not belong to the set
$\textbf{K}(5)$. For the values 5, 7 the compatibility condition
holds automatically and as a consequence the coefficients $c_5$,
$c_7$ are arbitrary ones. The two-parametric power expansion that
corresponds to the reduced solutions \eqref{4.5} (at $c_{-1}=2$) is
\begin{equation*}\begin{gathered}
\label{}w(z)=\frac{2}{{z}}+2\,{z}^{-1}+ \left( {\frac
{1}{72}}+{\frac {1}{144}}\,\alpha \right) {
z}^{4}+c_{{5}}{z}^{5}+c_{{7}}{z}^{7}+
\\+ \left( {\frac {17}{370656}}+{
\frac {59}{1482624}}\,\alpha+{\frac {25}{2965248}}\,{\alpha}^{2}
 \right) {z}^{9}+c_{{5}} \left( {\frac {19}{3780}}+{\frac {1}{432}}\,
\alpha \right) {z}^{10}+
\\+{\frac {11}{78}}\,{c_{{5}}}^{2}{z}^{11}+c_{{7}
} \left( {\frac {127}{35280}}+{\frac {1}{576}}\,\alpha \right)
{z}^{12}+...,
\end{gathered}\end{equation*}
where $c_i\equiv c_i^{(3)},\,i=5,7$. Obtained expansions converge
for small $|z|$. The exponential additions for the expansions
\eqref{4.15}, \eqref{4.18} do not exist. The reduced equation
\eqref{4.1} does not have non-power solutions.

\section{Power expansions, corresponding to the edge $\Gamma_2^{(1)}$ at $\alpha \neq 0$.}

The edge $\Gamma_2^{(1)}$ is characterized by the reduced equation
\begin{equation}
\label{5.1}\hat{f}_2^{(1)} (z,w) \stackrel{def}{=} 6\,w^5 - zw=0
\end{equation}
and the normal cone
$U^{(1)}_2=\{\lambda(4,1)=4\lambda(1,1/4),\,\lambda>0\}$. It means
that $r=1/4$, $\omega=1$, $z\rightarrow  \, \infty$ and the solution
of the reduced equation is $y=c_{1/4}z^{1/4}$. Substitution this
expression into \eqref{5.1} and cancelation the result by $z^{5/4}$
yields the determining equation $ c_{1/4}(6c_{1/4}^4-1)=0$. Thus we
have four families of power asymptotics
\begin{equation}
\label{5.2}\mathcal{F}_2^{(1)}1:\,\,\,\, w=
c_{1/4}^{(1)}\,z^{1/4},\,\quad\,c_{1/4}^{(1)}=\left(\frac{1}{6}\right)^{1/4};
\end{equation}
\begin{equation}
\label{5.3}\mathcal{F}_2^{(1)}2:\,\,\,\, w=c_{1/4}^{(2)}\,
z^{1/4},\,\quad\,c_{1/4}^{(2)}\,=-\left(\frac {1}{6}\right)^{1/4};
\end{equation}
\begin{equation}
\label{5.4}\mathcal{F}_2^{(1)}3:\,\,\,\,w=c_{1/4}^{(3)}\,z^{1/4},\,\quad\,c_{1/4}^{(3)}\,=i\left(\frac
{1}{6}\right)^{1/4};
\end{equation}
\begin{equation}
\label{5.5}\mathcal{F}_2^{(1)}4:\,\,\,\, w=
c_{1/4}^{(4)}\,z^{1/4},\,\quad\,c_{1/4}^{(4)}=-i\left(\frac{1}{6}\right)^{1/4}.
\end{equation}
Since the reduced equation \eqref{5.1} is algebraic and the roots of
the determining equation are simple then asymptotics do not have
proper (and consequently  critical) numbers and $\nu(k)\equiv
const\neq0$. The shifted carrier of the power asymptotics
\eqref{5.2} -- \eqref{5.5} gives a vector $B=\left(1/4,-1\right)$.
Points belonging to the lattice generated by vectors $B_{2}$, $B$
are the following $Q=(q_1,q_2) =m(1,\,1) +l(1/4,\,-1)
=\left(m+l/4,\,\,m-l \right)$ where $m$, $l$ are whole numbers. At
the line $q_2=-1$ we have $q_2=-1$ and $q_1=-1+5/4l$. Taking into
consideration that the cone of the problem is
$\mathcal{K}=\{k<1/4\}$ we find the set $\mathbf{K}$
\begin{equation}
\label{5.6}\mathbf{K}=\{k=-1-5l/4,\,\,\, l\in \mathbb{N}\cup\{0\}
\},
\end{equation}
The expansions of the solutions can be written as
\begin{equation}
\label{5.7}G_2^{(1)} n:\,\,\,\, w(z)=\varphi^{(n)}(z)=c^{(n)}_{1/4}
z^{1/4} + \sum^{\infty}_{l=0} c^{(n)}_{-1-5l/4}\, z^{-1-5l/4}.
\end{equation}
In this expression coefficients $c^{(n)}_{-1-5l/4}$ can be
sequentially computed. The calculation of the coefficients $c_{-1}$
yields $c_{-1}=\alpha/4$. Taking into account four terms, the
expansions are
\begin{equation*}
\begin{gathered}
\label{}w(z)=\varphi^{(n)}(z)=c_{{1/4}}^{(n)}\,{z^{1/4}}+\frac14\,{\frac
{\alpha}{z}}-{\frac {5}{16}}\,{ \frac {(c_{{1/4}}^{(n)})^{3} \left(
1+3\,{\alpha}^{2} \right) }{{z}^{9/4}}}+
\\+{\frac {5}{128}}\,{\frac
{(c_{{1/4}}^{(n)})^{2}\alpha\, \left( 29+24\,{ \alpha}^{2}
\right)}{{z}^{7/2}}}+...
\end{gathered}
\end{equation*}
It is likely that the obtained expansions diverge. Non-power
asymptotics do not correspond to the edge $\Gamma^{(1)}_2$ but it
generates exponential additions which will be computed later.

\section{Power expansions, corresponding to the edge $\Gamma_3^{(1)}$ at $\alpha \neq 0$.}

The edge $\Gamma_3^{(1)}$ is characterized by the reduced equation
\begin{equation}
\label{6.1}\hat{f}_3^{(1)} (z,w) \stackrel{def}{=} - zw - \alpha=0
\end{equation}
and the normal cone $U^{(1)}_3=\{\lambda(1,-1),\,\lambda>0\}$. In
this case $r=-1$, $\omega=1$, $z\longrightarrow \, \infty$ and power
asymptotic can be presented in the form
\begin{equation}
\label{6.2}\mathcal{F}_3^{(1)}1:\,\,\,\, w= \frac
{c_{-1}}{z},\,\quad\,c_{-1}=-\alpha.
\end{equation}
As the equation \eqref{6.1} is algebraic, then its solutions do not
have critical numbers and
\begin{equation}
\label{6.3}\nu(k)=z^{-1}\frac{\delta f_3^{(1)}}{\delta y}=-1.
\end{equation}
The cone of the problem is $\mathcal{K}=\{k<-1\}$. The shifted
carrier of the power asymptotic \eqref{6.2} gives the vector
$(-1,-1)=-B_{2}$, which belongs to the lattice generated by the
carrier of the studied equation. That is why
\begin{equation}
\label{6.4}\mathbf{K}=\{k=-1-5m,\,\,\, m\in \mathbb{N} \}.
\end{equation}
So we have determined the power expansion corresponding to the
asymptotic \eqref{6.2}
\begin{equation}
\label{6.5}G_3^{(1)}:\,\,\,\, w(z)=\psi(z)=\frac {c_{-1}}{z} +
\sum^{\infty}_{m=1} c_{-1-5m}\, z^{-1-5m}.
\end{equation}
In this expression all coefficients can be sequentially found. Again
taking into account four terms, it can be rewritten as
\begin{equation*}
\begin{gathered}
\label{}w(z)=\psi(z)=-{\frac {\alpha}{z}}-6\,{\frac
{c}{{z}^{6}}}-12\,{\frac {c \left( 15\,
{\alpha}^{4}-295\,{\alpha}^{2}+1512 \right) }{{z}^{11}}}- \\
\\ -216\,{\frac { c \left(35\,{\alpha}^{8}-2090\,{\alpha}^{6}+49755\,{\alpha}^{4}-
528180\,{\alpha}^{2}+2018016 \right) }{{z}^{16}}}+...,
\end{gathered}
\end{equation*}
where $c=\alpha(\alpha-1)(\alpha-2)(\alpha+1)(\alpha+2)$ and all
other not written out coefficients are proportional to this factor.
Unless $c=0$ the expansion \eqref{6.5} seems to be divergent one for
all $|z|^{-1}\neq0$.

The reduced equation \eqref{6.1} does not have non-power solutions.
The edge $\Gamma^{(1)}_3$ defines exponential additions which will
be found below.

\section{Power expansions, corresponding to the edge $\Gamma_4^{(1)}$ at $\alpha \neq 0$.}

The edge $\Gamma_4^{(1)}$ defines the following reduced equation
\begin{equation}
\label{7.1}\hat{f}_4^{(1)} (z,w)\stackrel{def}{=}w_{zzzz} - \alpha=0
\end{equation}
and the normal cone $U^{(1)}_4 =\{-\lambda(1,4),\,\,\lambda>0\}$.
Thus $\omega=-1$, i.e. $z\rightarrow 0$ è $r=4$ and we have the
unique family of power asymptotics
\begin{equation}
\label{7.2}\mathcal{F}_4^{(1)}1:
\,\,\,w=c_4z^4,\,\,\,c_4=\frac{\alpha}{24}.
\end{equation}
Let us find the critical numbers. The first variation of the
equation \eqref{7.1} is
\begin{equation}
\label{7.3}\frac{\delta \hat{f}^{(1)}_4}{\delta w}
=\frac{d^4}{dz^4}.
\end{equation}
The proper numbers are $k_1=0,\,\,\, k_2=1,\,\,\, k_3=2,\,\,\,
k_4=3$. None of them belongs to the cone of the problem $
\mathcal{K}=\{k>4\}$. Consequently in this case there are no
critical numbers. The shifted carrier of \eqref{7.2} is equal to the
vector $B_1$ and the lattice generated by the vectors $B_1$, $B_2$
corresponds to the power asymptotic \eqref{7.2}. Because of this the
set $\textbf{K}$ is
\begin{equation}
\label{7.4}\textbf{K}=\{k=4+5m,\,\,\,m\in \mathbb{N}\}.
\end{equation}
The power expansion can be written as
\begin{equation}
\label{7.5}w(z)=z^4\left(\frac{\alpha}{24}+\sum^{\infty}_{m=1}
c_{4+5m}\,z^{5m}\right).
\end{equation}
The coefficient $c_{4+5m}$ can be uniquely computed. The first three
terms of the found expansion $G_4^{(1)}1$ are
\begin{equation*}
\label{}w(z)=\alpha\left(\frac1{24}\, {z}^{4}+{\frac
{1}{72576}}\,{z}^{9}+( {\frac {1}{1743565824}}+{\frac
{5}{5930496}}\,{\alpha}^{ 2} ) {z}^{14}+...\right).
\end{equation*}
The expansion \eqref{7.5} can be regarded as the special case of the
expansion \eqref{3.10} at $c_0=c_1=c_2=c_3=0$. It converge for small
$|z|$. The existence, the uniqueness and the analyticity of such
expansion follow from the above mentioned Caushy theorem.

The edge $\Gamma^{(1)}_4$ does not define exponential additions and
non-power asymptotics.

\section{General properties of the equation \eqref{1.3} at $\alpha = 0$.}

If $\alpha=0$ the fourth-order analogue to the second Painlev\'{e}
equation is
\begin{equation}
\label{8.1}\tilde{f}(z,w)\stackrel{def}{=}w_{zzzz} - 10w^2w_{zz} -
10ww^2_z + 6 w^5  - zw =0.
\end{equation}
The carrier $S(\tilde{f})$ of this equation consists of the four
points: $Q_1=(-4,1),\,\,Q_2=(0,5),\,\,Q_3=(1,1)$ è $Q_4=(-2,3)$.
Their convex hull is the triangle with the apexes
$\Gamma^{(0)}_j=Q_j$ $(j=1,2,3)$ and edges
$\Gamma^{(1)}_1=\{Q_1,Q_2\},\,\,\Gamma^{(1)}_2=\{Q_2,Q_3\},\,\,\Gamma^{(1)}_1=\{Q_3,Q_1\}$
(fig. 3).
\begin{figure}[h]
 \centerline{\epsfig{file=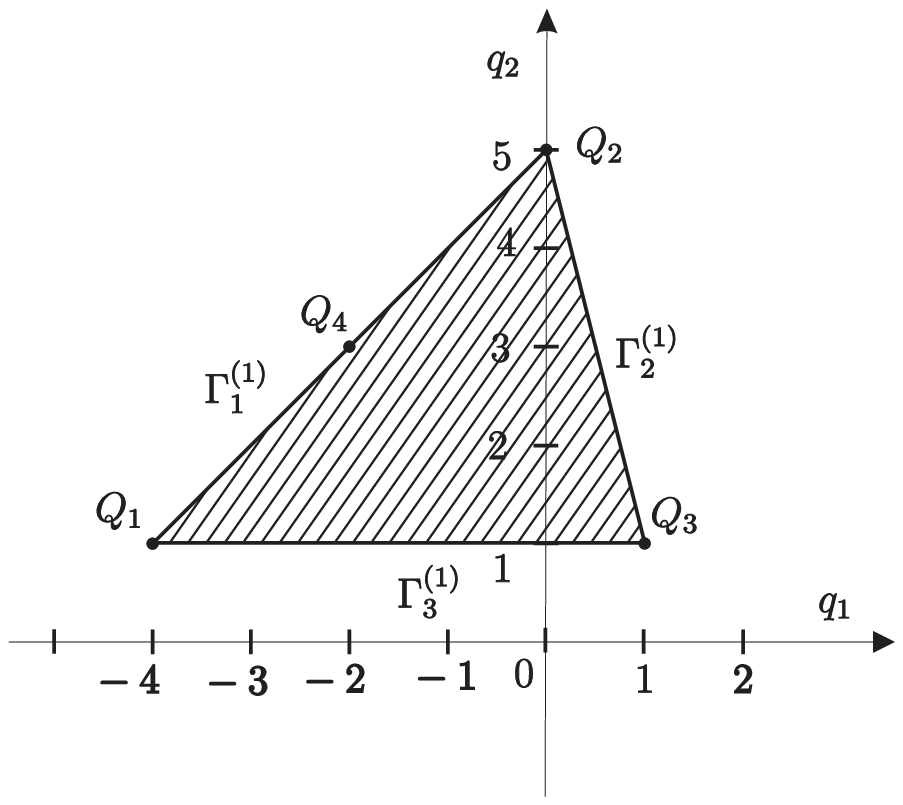,width=100mm}}
 \caption{}\label{fig:z_post}
\end{figure}

The external normals $N_j$ to the edges $\Gamma^{(1)}_j$ are the
vectors $N_1=(-1,1),\,\,N_2=(4,1),\,\,N_3=(0,-1)$. Normal cones
$U_j^{(1)}$ of the edges $\Gamma_j^{(1)}$ can be written as
\begin{equation}
\label{8.2}U_j^{(1)} =\{P=\lambda N_j,\,\,\, \lambda>0\},\,\,\,
j=1,2,3.
\end{equation}
Normal cones $U_j^{(0)}$ of the apexes $\Gamma_j^{(0)}$ and the
normal cones \eqref{8.2} are presented at the fig. 4. The shifted
carrier of the equation \eqref{8.1} lies in the lattice with the
basis $B_3=(-3,2)$ and $B_4=(-1,4)$.
\begin{figure}[h]
 \centerline{\epsfig{file=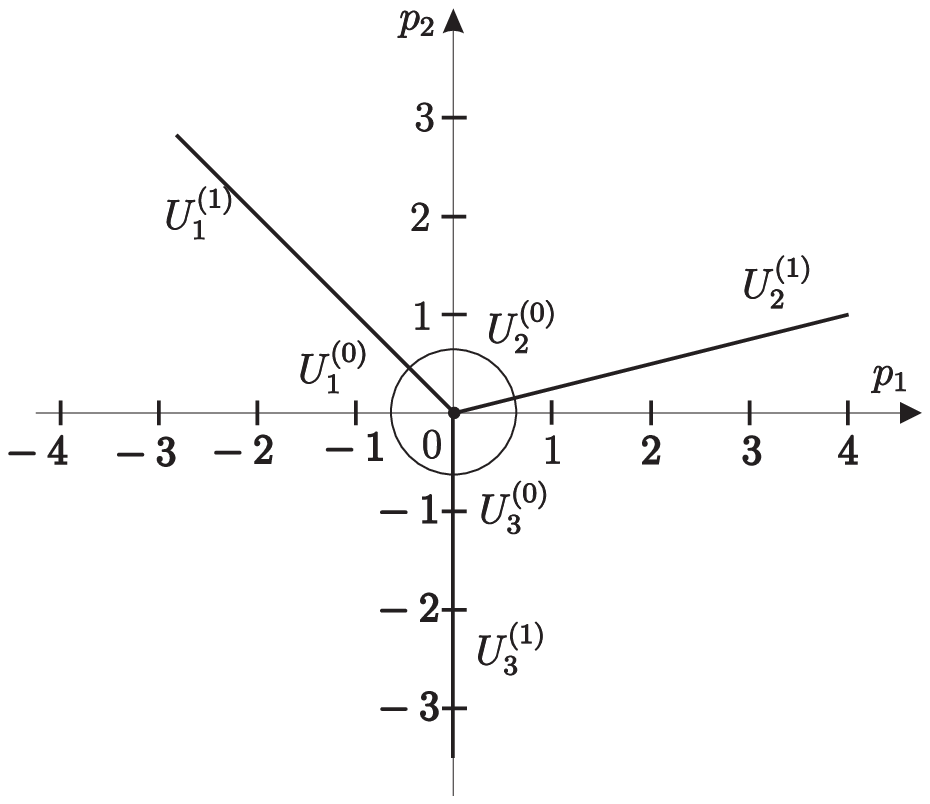,width=100mm}}
 \caption{}\label{fig:z_post}
\end{figure}

Further we will study the reduced equations corresponding to the
apexes $\Gamma_j^{(0)}$ and to the edges $\Gamma_j^{(1)}$
$(j=1,2,3)$. Again the reduced equations
\begin{equation}
\label{8.3}\hat{\tilde{f}}_2^{(0)}(z,w)\stackrel{def}{=}6w^5=0,\,\,\hat{\tilde{f}}_3^{(0)}(z,w)\stackrel{def}{=}-zw=0,
\end{equation}
which are defined by the apexes $\Gamma_2^{(0)}$, $\Gamma_3^{(0)}$
accordingly are the algebraic ones. Because of this they do not have
non-trivial power or non-power solutions.

\section{Power expansions, corresponding to the apex $\Gamma_1^{(0)}$ and to the edge $\Gamma^{(1)}_1$ at $\alpha =0$.}

The apex $\Gamma_1^{(0)}$ is characterized by the following reduced
equation
\begin{equation}
\label{9.1}\hat{\tilde{f}}_1^{(0)}(z,w)\stackrel{def}{=} w_{zzzz}=0.
\end{equation}
This case is similar to the one discussed in the section 3.
Consequently there are four families of power expansions:
four-parametric family $\tilde{G}_1^{(0)}1$, tree-parametric family
$\tilde{G}_1^{(0)}2$, two-parametric family $\tilde{G}_1^{(0)}3$ and
one-parametric family $\tilde{G}_1^{(0)}4$.

Examining by analogy with the section 4 the reduced equation which
corresponds to the edge $\Gamma^{(1)}_1$
\begin{equation}
\label{9.2}\hat{\tilde{f}}^{(1)}_1 (z,w) \stackrel{def}{=}w_{zzzz}
-10\,w^2\,w_{zz} -10\,w\,w_{z}^2+6\,w^5=0,
\end{equation}
we get two tree-parametric families $\tilde{G}_1^{(1)}1$,
$\tilde{G}_1^{(1)}2$ and two two-parametric families
$\tilde{G}_1^{(1)}3$, $\tilde{G}_1^{(1)}4$ of power expansions.

The wave in these designations means that in the similar expressions
for the equation \eqref{1.3} it is provided that $\alpha=0$.

\section{Power expansions, corresponding to the edge $\Gamma_2^{(1)}$ at $\alpha =0$.}

The edge $\Gamma^{(1)}_2$ defines the reduced equation which can be
written as
\begin{equation}
\label{10.1}\hat{\tilde{f}}_2^{(1)} (z,w) \stackrel{def}{=} 6\,w^5 -
zw=0.
\end{equation}
Following the analysis of the section 5 we get that
$r=1/4,\,\omega=1$, $z\rightarrow\infty$ and that there are four
families of power asymptotics: $\tilde{\mathcal{F}}_2^{(1)}1$
\eqref{5.2}, $\tilde{\mathcal{F}}_2^{(1)}2$ \eqref{5.3},
$\tilde{\mathcal{F}}_2^{(1)}3$ \eqref{5.4},
$\tilde{\mathcal{F}}_2^{(1)}4$ \eqref{5.5}. The shifted carrier of
the reduced solutions is the vector $(1/4,-1)$. Together with the
vectors $B_3$, $B_4$ it generates the lattice with the basis
$(1/4,-1)$ and $(-3,2)$. The points of this lattice are
$Q=(q_1,q_2)=m(1/4,-1)+l(-3,2)=(m/4-3l,-m+2l)$. At the line $q_2=-1$
we have $q_1=1/4+l/2-3l=1/4-5l/2$. As the cone of the problem is
$\mathcal{K}=\{k<1/4\}$, then the set
\begin{equation}
\label{10.2}\textbf{K}=\{k=1/4-5l/2,\,\,\,l\in \mathbb{N}\}
\end{equation}
differs from the similar in the section 5. Hence the power
expansions can be presented in the form
\begin{equation}
\label{10.3}\tilde{G}_2^{(1)} n:\,\,\,\,
w(z)=\varphi^{(n)}(z)=c^{(n)}_{1/4} z^{1/4} + \sum^{\infty}_{l=1}
c^{(n)}_{1/4-5l/2}\, z^{1/4-5l/2}.
\end{equation}
All coefficients in the expression \eqref{10.3} can be sequently
computed. Taking into account three terms these expansions can be
rewritten as
\begin{equation*}
\begin{gathered}
\label{}w(z)=\varphi^{(n)}(z)=c_{{1/4}}^{(n)}z^{1/4}-{\frac
{5}{16}}\,{\frac {(c_{{1/4}}^{(n)})^{3}}{{z}^{9 /4}}}-{\frac
{297}{512}}\,\frac{c_{{1/4}}^{(n)}}{{z}^{{19/4}}}+...
\end{gathered}
\end{equation*}
It is likely that obtained expressions are divergent ones.

\section{Non-power expansions, corresponding to the edge $\Gamma_3^{(1)}$ in the case $\alpha =0$.}

The reduced equation which corresponds to the edge $\Gamma^{(1)}_3$
\begin{equation}
\label{11.1}\hat{\tilde{f}}_2^{(3)} (z,w) \stackrel{def}{=}
w_{zzzz}-zw=0
\end{equation}
does not possess solutions in the form $w=c_{r}z^{r},\,c_r\neq0$ (it
has a trivial solution $w\equiv 0$ ). The edge $\Gamma^{(1)}_3$
defines non-power asymptotics of the equation \eqref{8.1}. It is
horizontal, consequently in order to find the solutions of the
reduced equation \eqref{11.1} it is necessary to make the
logarithmic transformation
\begin{equation}
\label{11.2}y=\frac{d \ln w}{dz}.
\end{equation}
Hence derivatives of $w(z)$ are
\begin{equation*}
\begin{gathered}
\label{}w'=yw,\,\,\,w''=(y'+y^2)w,\,\,\,w'''=(y''+3yy'+y^3)w,\\
w''''=(y'''+4yy''+3(y')^2+6y^2y'+y^4)w.
\end{gathered}
\end{equation*}
After application of this transformation and after cancelation of
the result by $w$  the equation \eqref{11.1} can be rewritten as
\begin{equation}
\label{11.3}h(z,y)\stackrel{def}{=}y'''+4yy''+3(y')^2+6y^2y'+y^4-z=0.
\end{equation}
The carrier $S(h)$ of this equation consists of five points:
$M_1=(-3,1),\,M_2=(0,4),\,M_3=(1,0),\,M_4=(-2,2)$ è $M_5=(-1,3)$.
Their convex hull $\Gamma(h)$ is the triangle with the apexes
$\tilde{\Gamma}_j^{(0)}=M_j,\,\,\,j=1,2,3$ and the edges
$\tilde{\Gamma}_1^{(1)}=\{M_1,M_2\}\,\,\,\tilde{\Gamma}_2^{(1)}=\{M_2,M_3\}\,\,\,\tilde{\Gamma}_3^{(1)}=\{M_3,M_1\}$
(fig. 5).
\begin{figure}[h]
 \centerline{\epsfig{file=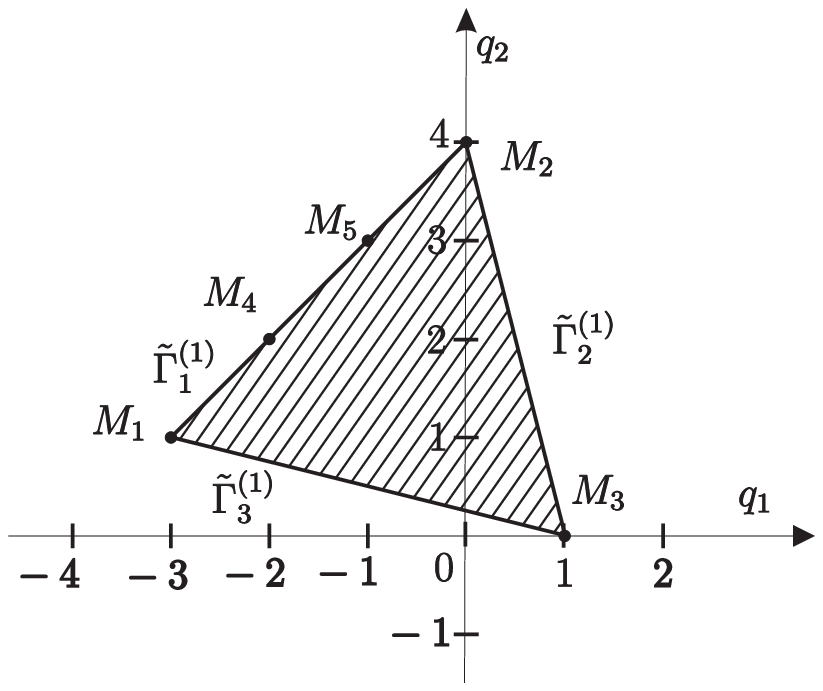,width=100mm}}
 \caption{}\label{fig:z_post}
\end{figure}

The external normals $\tilde{N}_j$ to the edges
$\tilde{\Gamma}_j^{(1)}$ are
$\tilde{N}_1=(-1,1),\,\tilde{N}_2=(4,1),\,\tilde{N}_3=(-1,-4)$. The
normal cones of the apexes and of the edges are presented at the
fig. 6. The shifted carrier of the equation \eqref{11.3} lies in the
lattice with the basis $B_5=(1,1),\,B_6=(1,-4)$. In this case the
cone of the problem is $p_1+p_2>0$. It intersects with the normal
cones $\tilde{U}_2^{(0)},\,\tilde{U}_3^{(0)},\,\tilde{U}_2^{(1)}$.
That is why later it is sufficient to study the reduced equations
which correspond to the bounds
$\tilde{\Gamma}_2^{(0)},\,\tilde{\Gamma}_3^{(0)},\,\tilde{\Gamma}_2^{(1)}$
only.
\begin{figure}[h]
 \centerline{\epsfig{file=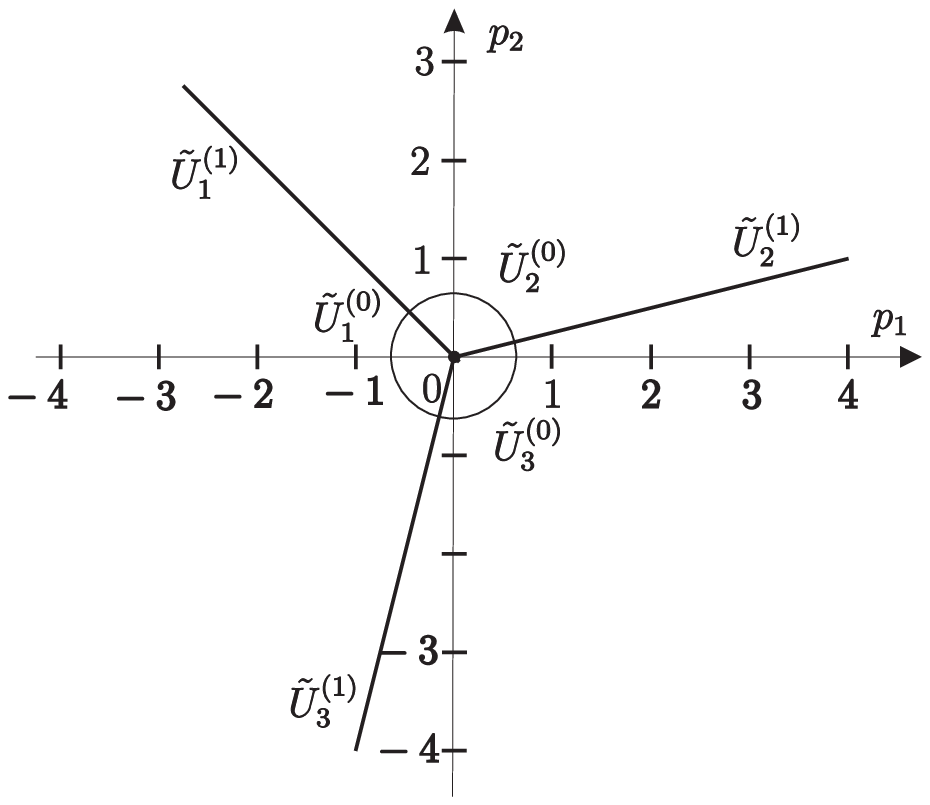,width=100mm}}
 \caption{}\label{fig:z_post}
\end{figure}
The first two of them are characterized by the algebraic reduced
equations $y^4=0,\,-z=0$, accordingly. They do not give suitable
solutions. The edge $\tilde{\Gamma}_2^{(1)}$ defines algebraic
reduced equation
\begin{equation}
\label{11.4}y^4-z=0,
\end{equation}
which has four suitable solutions $y^{(n)}(z)=c_{1/4}^{(n)} z^{1/4}$
where $c_{1/4}^{(n=\{1,2\})}=\pm1,\,c_{1/4}^{(n=\{3,4\})}=\pm i$ and
$r=1/4,\,\omega=1$. These solutions do not have critical numbers.
The shifted carrier of the equation \eqref{11.3} is the vector
$(1/4,-1)=1/4B_6$. Thus we have a lattice with the basis
$(1/4,-1),\,(1,1)$. Taking into account the fact that the cone of
the problem is $ \mathcal{K}=\{k<1/4\}$, we find the set
$\textbf{K}$
\begin{equation}
\label{11.5}\textbf{K}=\{k=1/4-5l/4,\,\,\,l\in \mathbb{N}\}.
\end{equation}
The equation \eqref{11.3} has solutions in the form of the series
\begin{equation}
\label{11.6}y^{(n)}(z)=c^{(n)}_{1/4} z^{1/4} + \sum^{\infty}_{l=1}
c^{(n)}_{1/4-5l/4}\, z^{1/4-5l/4},\,\,\,n=1,2,3,4,
\end{equation}
where the coefficients $c^{(n)}_{1/4-5l/4}$ are uniquely defined.
The coefficient $c^{(n)}_{-1}$ does not depend on $n$ and
$c^{(n)}_{-1}=-3/8$. Introducing the new variable $\eta=\ln w$ we
get
\begin{equation}
\label{11.7}\eta =\int ydz=
\frac45c^{(n)}_{1/4}z^{5/4}+C_0-\frac38\ln z+\sum^{\infty}
_{l=2}\frac{4c^{(n)}_{1/4-5l/4}}{5(1-l)}z^{5(1-l)/4}.
\end{equation}
In this expression $C_0$ is a constant of integration. Returning to
the variable $w(z)$ yields
\begin{equation}
\begin{gathered}
\label{11.8}w(z)=\frac{C_1}{z^{3/8}}\exp\left[\frac45c^{(n)}_{1/4}z^{5/4}+\sum^{\infty}
_{l=2}\frac{4c^{(n)}_{1/4-5l/4}}{5(1-l)}z^{5(1-l)/4}\right].
\end{gathered}
\end{equation}
Taking into account two terms of the series the obtained non-power
asymptotics of the equation \eqref{8.1} can be written as
\begin{equation*}
\begin{gathered}
\label{}w(z)=\frac{C_1}{z^{3/8}}\exp\left[\frac45c^{(n)}_{1/4}z^{5/4}+{\frac
{9}{32}}\,{\frac {1}{c^{(n)}_{1/4}{z}^{5/4} }}+{\frac
{45}{256}}\,\frac {1}{(c^{(n)}_{1/4})^{2}{z}^{5/2}}+... \right].
\end{gathered}
\end{equation*}
Providing that $w\rightarrow 0$ at $z\rightarrow\infty$, we should
take the number $n$ in \eqref{11.8} so that
\begin{equation}
\label{11.9}Re(c^{(n)}_{1/4}z^{5/4})<0.
\end{equation}
Consequently we have found four one-parametric non-power asymptotics
for the solutions of the equation \eqref{8.1}:
$\tilde{G}_3^{(1)}1,\,\tilde{G}_3^{(1)}2,\,\tilde{G}_3^{(1)}3,\,\tilde{G}_3^{(1)}4$.

\section{Exponential additions to the solutions.}

Let us find exponential additions to the obtained solutions. Since
the equations $\partial\hat{f}^{(d)}_{j}/\partial w^{\Pi(f)}=0$ do
not have solutions for $\forall (d=0,1; j=1,2,3,4)$ and the
condition $\Pi(\hat{f}^{(d)}_{j}) < \Pi(f)=4$ holds for the edges
$\Gamma^{(1)}_2$ and $\Gamma^{(1)}_3$ only (in the case
$\alpha\neq0$) then these very edges define exponential additions
which are to be computed. At $\alpha=0$ the condition
$\Pi(\hat{\tilde{f}}^{(d)}_{j}) < \Pi(\tilde{f})=4$ is true for the
edge $\Gamma^{(1)}_2$, consequently solutions corresponding to this
edge possess exponential additions which can be calculated as it
will be done for the edge $\Gamma^{(1)}_2$ of the equation
\eqref{1.3}.

\section{Exponential additions of the first level, corresponding to the edge $\Gamma^{(1)}_2$ at $\alpha\neq0$.}

To begin with we will look for the exponential additions of the
first level $u^{(n)}(z)$ to the expansions \eqref{5.7}, i.e. we will
look for the functions
\begin{equation}
\label{13.1} w(z)=\varphi^{(n)}(z) + u^{(n)}(z),\,\,\, n=1,2,3,4.
\end{equation}
The reduced equation for the addition $u^{(n)}(z)$ is a linear
equation
\begin{equation}
\label{13.2}M_{n}^{(1)}(z) u^{(n)}(z)=0,
\end{equation}
where $M_{n}^{(1)}(z)$ is the first variation at the solutions
$w(z)=\varphi^{(n)}(z)$. As long as
\begin{equation}
\label{13.3}\frac{\delta f}{\delta w} =\frac{d^4}{dz^4}
-10w^{2}\frac{d^2}{dz^2} -20ww_z \frac{d}{dz}- 20 w_{zz}w
-10w_z^{2}+30w^4-z,
\end{equation}
then
\begin{equation}
\label{13.4}M_{n}^{(1)}(z) =\frac{d^4}{dz^4}
 -10(\varphi^{(n)})^{2}\frac{d^2}{dz^2}
-20\varphi^{(n)}\varphi^{(n)}_z \frac
d{dz}-20\varphi^{(n)}_{zz}\varphi^{(n)} -10(\varphi^{(n)}_z)^{2}+
30(\varphi^{(n)})^4 -z.
\end{equation}
And equation \eqref{13.2} can be rewritten as
\begin{equation}
\begin{gathered} \label{13.5}\frac{d^4u^{(n)}}{dz^4}  -10
(\varphi{(n)})^{2} \frac{d^2u^{(n)}}{dz^2}-20
\varphi^{(n)}\varphi^{(n)} _z \frac{du^{(n)}}{dz}-\\
-[10(2\varphi^{(n)}_{zz}\varphi^{(n)} +(\varphi^{(n)}_z)^{2}
-3(\varphi^{(n)})^4) +z]\,u^{(n)}=0,\,\,\,
 n=1,2,3,4.
\end{gathered}
\end{equation}
Let us make the change of the variables
\begin{equation}
\label{13.6}\zeta^{(n)}(z)=\frac{d \ln u^{(n)}(z)}{dz}.
\end{equation}
Computing the derivatives of the function $u^{(n)}(z)$ yields
\begin{equation*} \label{}\frac{du^{(n)}}{dz} =\zeta^{(n)} u^{(n)},
\,\,\quad\, \frac{d^2u^{(n)}}{dz^2 }=\zeta _z^{(n)} u^{(n)} + (\zeta
^{(n)})^2 u^{(n)},
\end{equation*}
\begin{equation*}
\label{}\frac{d^3u^{(n)}}{dz^3} =\zeta^{(n)}_{zz} u^{(n)}+ 3 \zeta
^{(n)} \zeta_z^{(n)} u^{(n)} + (\zeta^{(n)})^3 u^{(n)},
\end{equation*}
\begin{equation*}
\label{}\frac{d^4u^{(n)}}{dz^4} =\zeta^{(n)}_{zzz} u^{(n)}+ 4 \zeta
^{(n)}\zeta^{(n)}_{zz} u^{(n)} + 3 (\zeta_z^{(n)})^2  u^{(n)} + 6
(\zeta^{(n)})^2 \zeta _z^{(n)} u^{(n)} + (\zeta ^{(n)})^4 u^{(n)}.
\end{equation*}
Substituting these expressions into the equation \eqref{13.5} and
setting to zero the factor at $u^{(n)}(z)$ we get the equation
\begin{equation}
\begin{gathered}
\label{13.7} \zeta^{(n)}_{zzz} + 4\zeta^{(n)}\zeta^{(n)}_{zz}
+3(\zeta^{(n)}_z)^2 +6 (\zeta^{(n)})^2
\zeta^{(n)}_z +  (\zeta^{(n)})^4 -20\varphi_z^{(n)}\varphi_{zz}^{(n)} +\\
  -10(\varphi^{(n)})^2( \zeta^{(n)}_z + (\zeta^{(n)})^2)
-20\varphi^{(n)}\varphi_z^{(n)} \zeta^{(n)}
-10(\varphi_z^{(n)})^2+30 (\varphi^{(n)})^4 -z=0.
\end{gathered}
\end{equation}
It is essential to find power expansions of its solutions. The
carrier of the equation \eqref{13.7} consists of the points
\begin{equation}
\begin{gathered}
\label{13.8}Q_{1,1} =(-3,1),\,\,\, Q_{1,2}=(-2,2),\,\,\, Q_{1,3}=(-1,3), \\
Q_{1,4}=(0,4), \,\,\, Q_{1,5}=(1,0), \,\,\,Q_{2,\,k}=\left(\frac14-\frac{5}4k,0\right),\,\,\, \\
Q_{3,\,k}=\left(-\frac12-\frac{5}4k,1\right),\,\,
Q_{4,\,k}=\left(\frac12-\frac{5}4k,2\right),\,\,
Q_{5,\,k}=\left(-\frac32-\frac{5}4k,0\right),\\
Q_{6,\,k}=\left(1-\frac{5}4k,0\right),\,\, \,\,\ k \in
\mathbb{N}\cup\{0\},
\end{gathered}
\end{equation}
where the doubled index in the points numeration is introduced for
the notational convenience. The convex closing of the points
\eqref{13.8} is the half-string presented at the fig. 7.
\begin{figure}[h]
 \centerline{\epsfig{file=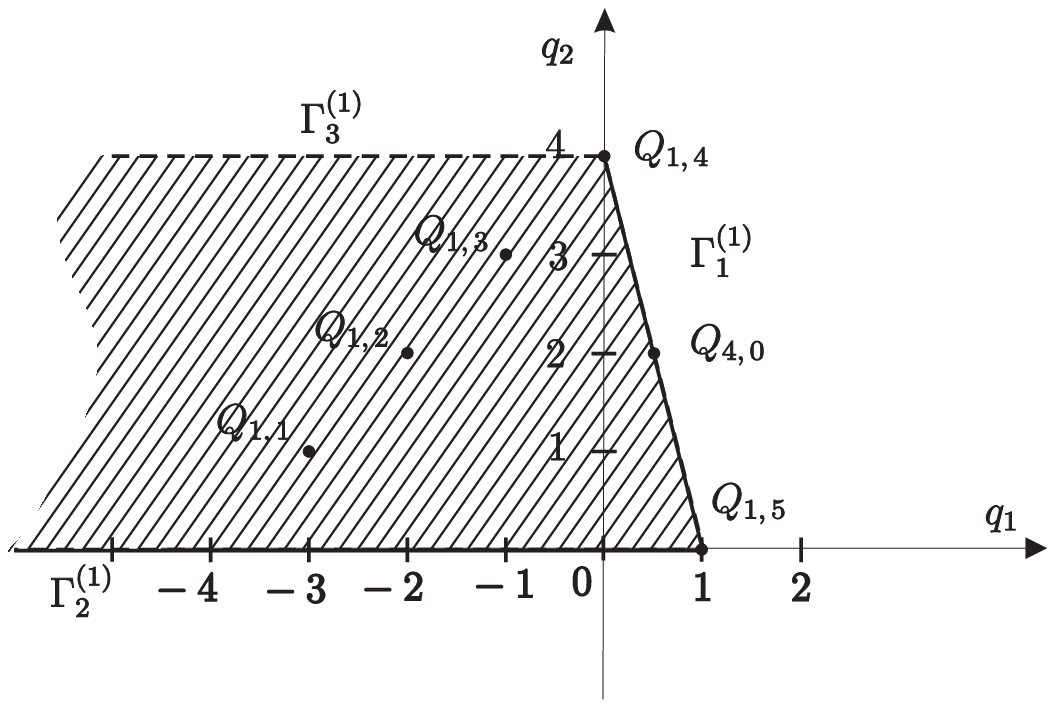,width=110mm}}
 \caption{}\label{fig:z_post}
\end{figure}
The periphery of the half-string contains two apexes
$Q_{1,\,4},Q_{1,\,5}$ and three edges $\Gamma^{(1)}_j\,\,(j=1,2,3)$
with the normal vectors $N_1=(4,1),\,\, N_2=(0,-1),\,\, N_3=(0,1)$.
Suitable expansions can be obtained examining the reduced equation
corresponding to the edge $\Gamma^{(1)}_1$
\begin{equation}
\label{13.9}h_1^{(1)}(z,\zeta^{(n)}) \stackrel{def}{=}
(\zeta^{(n)})^4 -10(c_{1/4}^{(n)})^2 (\zeta^{(n)})^2z^{\frac12} +
(30(c_{1/4}^{(n)})^4-1)z=0.
\end{equation}
It has sixteen solutions
\begin{equation}\begin{gathered}
\label{13.10}\zeta^{(n,\,l)}(z) =g_{1/4}^{(n,\,l)}z^{1/4},\,\,\,
n,l=1,2,3,4\,\,\,,
\end{gathered}
\end{equation}
where
\begin{equation*}\begin{gathered}
\label{}g_{1/4}^{(n;\,\,\,l=\{1,2,3,4\})}=\pm\left(\,\,\,5(c_{1/4}^{(n)})^2\pm\sqrt{1-5(c_{1/4}^{(n)})^4}\,\,\,\right)^{1/2}.
\end{gathered}\end{equation*}
Taking into account the fact that $(c_{1/4}^{(n)})^4=1/6$, we get
\begin{equation*}\begin{gathered}
\label{}g_{1/4}^{(n;\,\,\,l=\{1,2\})}=\pm6^{1/4},n=1,2\,\,\,;\,g_{1/4}^{(n;\,\,\,l=\{3,4\})}=\pm(8/3)^{1/4},n=1,2;\\
g_{1/4}^{(n;\,\,\,l=\{1,2\})}=\pm6^{1/4}i,n=3,4\,\,\,;\,g_{1/4}^{(n;\,\,\,l=\{3,4\})}=\pm(8/3)^{1/4}i,n=3,4.
\end{gathered}\end{equation*}

The reduced equation is an algebraic one that is why it does not
have critical numbers. Now we will look for the power expansions of
the equation \eqref{13.7} solutions which possess the power
asymptotic \eqref{13.10}. The shifted carrier of the equation
\eqref{13.7} lies in the lattice generated by the vectors
$B_7=\left(5/4,0\right),\,\, B_8=(1,1)$. The shifted carrier of the
reduced solutions \eqref{13.10} gives the vector $B_9=\left(-1/4,1
\right)$. Taking into consideration the correlation
$B_8-B_9=\left(5/4,0\right)= B_7$ we get that the vector $B_9$
belongs to the lattice with the basis $B_7,\,B_8$. The points of
this lattice are the following
\begin{equation*}
Q=(q_1,q_2)=k(1,1)
+m\left(\frac54,0\right)=\left(k+\frac{5m}4,k\right).
\end{equation*}
At the line $q_2=-1$ we have $k=-1$ and hence $q_1=-1+5m/4$. Since
the cone of the problem is $\mathcal{K}=\left\{k<\frac14\right\}$,
then the set $\mathbf{K}$ can be presented in the form
\begin{equation}
\label{13.11}\mathbf{K}=\left\{\frac{1-5k}4,k\in\mathbb{N}\right\}.
\end{equation}
Power expansions of the equation \eqref{13.7} solutions are
\begin{equation}\begin{gathered}
\label{13.12}\zeta^{(n,\,l)}(z)=g^{(n,\,l)}_{1/4} z^{1/4}
+\sum_{k\,\in\,\,\mathbb{N}}g^{(n,\,l)}_{(1-5k)/4}\,z^{(1-5k)/4},\,\,\,
 n,l=1,2,3,4.
\end{gathered}\end{equation}
Using the values of the coefficients $c_{1/4}^{(n)}$, $c_{-1}^{(n)}$
and $g^{(n,\,l)}_{1/4}$ we can compute the coefficient
$g^{(n,\,l)}_{-1}$.
\begin{equation*}
\begin{gathered}
\label{}
g_{-1}^{(n,\,l)}=-\frac38,n=1,2,3,4,\,l=1,2;\,\,\,g_{-1}^{(n,\,l)}=-\frac38+\frac54\alpha,(n,\,l)=\{(1,3),(2,4),(3,3),(4,4)\};\,\,\,\\
g_{-1}^{(n,\,l)}=-\frac38-\frac54\alpha,(n,\,l)=\{(1,4),(2,3),(3,4),(4,3)\}.
\end{gathered}
\end{equation*}
In view of the transformation \eqref{13.6} the additions
$u^{(n,\,l)}(z)$ can be found. They are
\begin{equation}
\label{13.13}u^{(n,\,l)}(z) = C_1 \exp \int \zeta^{(n,\,l)}(z) dz.
\end{equation}
Where from we get
\begin{equation}
\begin{gathered}
\label{13.14}u^{(n,\,l)}(z)=C_1\,z^{g^{(n,\,l)}_{-1}}\, \exp
\left[\frac45\, g^{(n,\,l)}_{1/4}\,z^{5/4} + \sum^{\infty}_{k=2}
\frac{4}{5(1-k)} g^{(n,\,l)}_{(1-5k)/4}
z^{5(1-k)/4}\right],\\
\,\, n,l=1,2,3,4.
\end{gathered}
\end{equation}
Here $C_1$ and later $C_2$ and $C_3$ are the constants of
integration. The additions $u^{(n,\,l)}(z)$ are exponentially small
at $z\rightarrow \infty$ in those sectors of the complex plane $z$,
where
\begin{equation}
\label{13.15}Re \left[g^{(n,\,l)}_{1/4}\, z^{5/4}\right]<0.
\end{equation}

Consequently for each expansion $G^{(1)}_{3}n$ we have found four
one-parametric families of additions $G_2^{(1)}n\, G^{(1)}_1 l$
($n,l=1,2,3,4$).

\section{Exponential additions of the second level, corresponding to the edge $\Gamma^{(1)}_2$ at $\alpha\neq0$.}

In this section we will compute the exponential additions of the
second level $v^{(n,\,l)}(z)$, i.e. the additions to the solutions
$\zeta^{(n,\,l)}(z)$. The reduced equation to the addition
$v^{(n,\,l)}(z)$ is
\begin{equation}
\label{14.1}M_{n,\,l}^{(2)} (z) v^{(n,\,l)}(z)=0,
\end{equation}
where an operator $M_{n,\,l}^{(2)}(z)$ is the first variation of
\eqref{13.7} at the solutions $\zeta^{(n,\,l)}(z)$. The equation
\eqref{14.1} can be rewritten as
\begin{equation}
\begin{gathered}
\label{14.2}\frac{d^3v}{dz^3} + 4\zeta \frac{d^2v}{dz^2} +
2(3\zeta_z + 3\zeta^{2}-5\varphi^2)\frac{dv}{dz} +  4(\zeta_{zz}
+3\zeta\zeta_z + \zeta^3 - 5\varphi^2\zeta - 5\varphi\varphi_z)v=0.
\end{gathered}
\end{equation}
Here and later $\varphi\stackrel{def}{=}\varphi^{(n)}$,
$v\stackrel{def}{=} v^{(n,\,l)}$ è $\zeta\stackrel{def}{=}
\zeta^{(n,\,l)}$. Making the transformation of the variables
\begin{equation}
\label{14.3}\frac{d \ln v}{dz}=\xi,
\end{equation}
we get
\begin{equation*}
\label{}\frac{dv}{dz}=\xi v,\,\,\quad\, \frac{d^2v}{dz^2} =\xi_z
v+\xi^2 v,\,\,\quad\, \frac{d^3 v}{dz^3}=\xi_{zz} v + 3\xi\xi_{z}
v+\zeta^3 v.
\end{equation*}
And the equation \eqref{14.2} transferees to
\begin{equation}\begin{gathered}
\label{14.4}\xi_{zz} + 3\, \xi\,\xi_{z} + \xi^3
+4\zeta(\xi_z+\xi^2)+6(\zeta_z+\zeta^2)\xi -10\,\varphi^2\xi+
4\zeta_{zz} + 12 \zeta\zeta_z  +
\\+4\,\zeta^3  -20\, \varphi^2\,\zeta
-20\,\varphi\,\varphi_z=0.
\end{gathered}\end{equation}
The carrier of this equation is defined by the following set of
points
\begin{equation}
\begin{gathered}
\label{14.5}M_{1,\,1}=(-2,1),\,\,\, M_{1,\,2}=(-1,2),\,\,\,
M_{1,\,3}=(0,3),\,\,\,
M_{2,\,k}=\left(-\frac34-\frac54k,\,1\right),\,\,\,\\
M_{3,\,k}=\left(\frac14-\frac54k,2\right),\,\,\,M_{4,\,k}=\left(\frac12-\frac54k,1\right),\,\,\,M_{5,\,k}=\left(-\frac74-\frac54k,\,0\right),\,\,\,\\
M_{6,\,k}=\left(-\frac12-\frac54k,\,0\right),\,\,\,M_{7,\,k}=\left(\frac34-\frac54k,\,0\right),\,\,
\, k \in \mathbb{N}\cup\{0\}.
\end{gathered}
\end{equation}
Their convex hull is the half-string similar to the one presented at
the fig. 7. Its edges  $\Gamma_3^{(1)}$, $\Gamma_2^{(1)}$ are the
rays starting out of the points $(0,\,3)$, $(3/4,\,0)$, accordingly,
and the edge $\Gamma_1^{(1)}$ connects the points $(3/4,\,0)$ è
$(0,\,3)$. Examining the edge $\Gamma_1^{(1)}$ which contains the
points $(0,\,3)$, $(1/4,\,2)$, $(1/2,\,1)$ and $(3/4,\,0)$ we can
find suitable solutions. The reduced equation corresponding to this
edge is
\begin{equation}
\begin{gathered}
\label{14.6}\xi^3 + 4g^{(n,\,l)}_{1/4}z^{1/4}\xi^2 +
2[3(g^{(n,\,l)}_{1/4})^2-5(c_{1/4}^{(n)})^2]z^{1/2}\xi +\\
+4g^{(n,\,l)}_{1/4}[(g^{(n,\,l)}_{1/4})^2 -
5(c_{1/4}^{(n)})^2]z^{3/4} =0.
\end{gathered}
\end{equation}
The solutions of the equation \eqref{14.6} can be presented in the
form
\begin{equation}
\begin{gathered}
\label{14.7}\xi^{(n,\,l,\,m)}(z)=r_{1/4}^{(n,\,l,\,m)}\,
z^{1/4},\,\,\,\, n,\,l=1,2,3,4;\,\,\,\, m=1,2,3,
\end{gathered}
\end{equation}
where $r\stackrel{def}{=}r_{1/4}^{(n,l,m)}$ are the roots of the
cubic equation
\begin{equation}
\begin{gathered}
\label{14.8} r^3 +4\,g_{1/4}^{(n,\,l)}r^2+
2\left(3\,(g_{1/4}^{(n,\,l)})^2 -5\,(c_{1/4}^{(n)})^2\right)\,r+\\
+4\,(g_{1/4}^{(n,\,l)})^{3} -20\,g_{1/4}^{(n,\,l)}\,(c_{1/4}^{(n)})^2=0.
\end{gathered}
\end{equation}
Solving this equation we get
\begin{equation*}
\begin{gathered}
\label{}r_{1/4}^{(n,\,l,\,1)}=-2\,g_{1/4}^{(n,\,l)},\,\quad\,
r_{1/4}^{(n,\,l,\,2)}=-g_{1/4}^{(n,\,l)}+\left({10\,(c_{1/4}^{(n)})^2-(g_{1/4}^{(n,\,l)})^2}\right)^{1/2}\\
r_{1/4}^{(n,\,l,\,3)}=-g_{1/4}^{(n,\,l)}-\left({10\,(c_{1/4}^{(n)})^2-(g_{1/4}^{(n,\,l)})^2}\right)^{1/2}.
\end{gathered}
\end{equation*}

The vectors $(1,\,1)$, $(5/4,\,0)$ are the basis of the lattice
which corresponds to the shifted carrier of the equation
\eqref{14.4}. The shifted carrier $(-1/4,1)$ of the reduced
solutions \eqref{14.7} belongs to this lattice. Consequently the set
$\mathbf{K}$ coincides with \eqref{13.11}. Power expansions of the
functions $\xi^{(n,\,l,\,m)}(z)$ are the following
\begin{equation}
\begin{gathered}
\label{14.9} \xi^{(n,\,l,\,m)}(z)=r_{1/4}^{(n,\,l,\,m)} z^{1/4}
+\sum_{k\,\in\,\,\mathbb{N}}r^{(n,\,l,\,m)}_{(1-5k)/4}\,z^{(1-5k)/4},\\
\,\quad\,n,l=1,2,3,4;\,\,\quad\, m=1,2,3.
\end{gathered}
\end{equation}
Calculation of the coefficients $r_{-1}^{(n,\,l,\,m)}$ yields
$r_{-1}^{(n,\,l,\,m)}=1/4$ at $n=1,2,3,4;\,\,\,l=1,2;\,\,\,m=1$. In
other cases $r_{-1}^{(n,\,l,\,m)}$ depend on the parameter $\alpha$.

So we have found the exponential additions $v^{(n,\,l,\,m)}(z)$ to
the solutions $\zeta^{(n,\,l)}(z)$
\begin{equation}
\begin{gathered}
\label{14.10}v^{(n,\,l,\,m)}(z)=C_2\,z^{r_{-1}^{(n,\,l,\,m)}}\, \exp
\left[\frac45\, r_{1/4}^{(n,\,l,\,m)}\,z^{5/4} + \sum^{\infty}_{k=2}
\frac{4}{5(1-k)} r^{(n,\,l,\,m)}_{(1-5k)/4}
z^{5(1-k)/4}\right]\\
\,\, n=1,2,3,4;\,\quad\,l=1,2,3,4;\,\quad\,m=1,2,3.
\end{gathered}
\end{equation}
They are exponentially small provided that
\begin{equation}
\label{14.11}Re
\left[r_{1/4}^{(n,\,l,\,m)}\,z^{5/4}\right]<0,\,\,z\rightarrow\infty.
\end{equation}

Thus, taking into account two-level additions $G_2^{(1)}n\,
G^{(1)}_1 l\,G^{(1)}_1 m$, we have $48$ expansions of the equation
\eqref{1.3} solutions ($n,l=1,2,3,4;\,\,\,m=1,2,3$).

\section{Exponential additions of the third level, corresponding to the edge $\Gamma^{(1)}_2$ at $\alpha\neq0$.}

In this section we will look for exponential additions of the third
level $y^{(n,\,l,\,m)}(z)$, i.e. we will look for additions to the
solutions $\xi^{(n,\,l,\,m)}(z)$. The reduced equation to the
addition $y^{(n,\,l,\,m)}(z)$ is the following
\begin{equation}
\label{15.1}M_{n,\,l,\,m}^{(3)} (z) y^{(n,\,l,\,m)}=0.
\end{equation}
Operator $M_{n,\,l,\,m}^{(3)}(z)$ can be found as the first
variation of \eqref{14.4} at the solutions $\xi^{(n,\,l,\,m)}(z)$
and then the equation \eqref{15.1} for the function
$y\stackrel{def}{=}y^{(n,\,l,\,m)}$ can be rewritten as
\begin{equation}
\begin{gathered}
\label{15.2}y_{zz} + (3\xi + 4\zeta)y_z +
(3\xi_z+3\xi^2+8\xi\zeta+6\zeta_z+6\zeta^2-10\varphi^2)y=0.
\end{gathered}
\end{equation}
Introducing the new variable $\eta$ by the rule
\begin{equation}
\label{15.3}\frac{d \ln y}{dz}=\eta
\end{equation}
we get that
\begin{equation*}
\label{}\frac{dy}{dz}=\eta y,\,\,\quad\, \frac{d^2y}{dz^2} =\eta_z
y+\eta^2 y.
\end{equation*}
Hence equation \eqref{15.2} transfers to the following
\begin{equation}\begin{gathered}
\label{15.4}\eta_{z} +\eta^2+ (3\xi +
4\zeta)\eta+3\xi_z+3\xi^2+8\xi\zeta+6\zeta_z+6\zeta^2-10\varphi^2=0.
\end{gathered}\end{equation}
The carrier of this equation is composed of the points
\begin{equation}
\begin{gathered}
\label{15.5}R_{1,\,1}=(-1,\,1),\,\,\, R_{1,\,2}=(0,\,2),\,\,\, R_{2,\,k}=(\frac14-\frac54k,\,1),\,\,\, R_{3,\,k}=\left(-\frac34-\frac54k,\,0\right),\\
R_{4,\,k}=\left(\frac12-\frac54k,\,0\right),\,\, \, k \in
\mathbb{N}\cup\{0\}.
\end{gathered}
\end{equation}
Closing the convex hull based on these points we obtain the
half-string similar to the one presented at fig. 7. Its edges
$\Gamma_3^{(1)}$ and $\Gamma_2^{(1)}$ are the rays starting out of
the points $(0,\,2)$ and $(1/2,\,0)$, accordingly. The edge
$\Gamma_1^{(1)}$ is limited by the points $(0,\,2)$ and $(1/2,\,0)$.
Now we will examine the reduced equation
\begin{equation}
\begin{gathered}
\label{15.6}\eta^2+
(3r_{1/4}^{(n,\,l,\,m)}+4g_{1/4}^{(n,\,l)})z^{1/4}\eta+
\left(3(r_{1/4}^{(n,\,l,\,m)})^2+8r_{1/4}^{(n,\,l,\,m)}g_{1/4}^{(n,\,l)}+\right.\\
\left. 6(g_{1/4}^{(n,\,l)})^2-10(c_{1/4}^{(n)})^2 \right)z^{1/2}=0,
\end{gathered}
\end{equation}
which corresponds to the edge $\Gamma_1^{(1)}$ (this edge contains
three points $(0,\,2)$, $(1/4,\,1)$ and $(1/2,\,0)$). The equation
\eqref{15.6} possesses the following solutions
\begin{equation}
\begin{gathered}
\label{15.7}\eta^{(n,\,l,\,m,\,p)}(z)=q^{(n,\,l,\,m,\,p)}_{1/4}\,
z^{1/4};\,\,\, n,l=1,2,3,4;\,\,\,\, m=1,2,3; \,\,\,\,p=1,2\,
\end{gathered}
\end{equation}
where $q\stackrel{def}{=}q^{(n,\,l,\,m,\,p)}_{1/4}$ are the
solutions of the quadratic equation
\begin{equation}
\begin{gathered}
\label{15.8}q^2+ (3r_{1/4}^{(n,\,l,\,m)}+4g_{1/4}^{(n,\,l)})q+
3(r_{1/4}^{(n,\,l,\,m)})^2+8r_{1/4}^{(n,\,l,\,m)}g_{1/4}^{(n,\,l)}+\\
6(g_{1/4}^{(n,\,l)})^2-10(c_{1/4}^{(n)})^2 =0.
\end{gathered}
\end{equation}
At fixed $n,\,l,\,m$ it has two solutions
\begin{equation*}
\begin{gathered}
\label{}
q_{1/4}^{(n,\,l,\,m,\,p=\{1,\,2\})}=-\frac32\,r_{1/4}^{(n,\,l,\,m)}-2\,g_{1/4}^{(n,\,l)}\pm\\
\pm\,\frac12\left({40\,(c_{1/4}^{(n)})^2-3\,(r_{1/4}^{(n,\,l,\,m)})^{2}-
-8\,{r_{1/4}^{(n,\,l,\,m)}}\,g_{1/4}^{(n,\,l)}-8\,
(g_{1/4}^{(n,\,l)})^{2}}\right)^{(1/2)}.
\end{gathered}
\end{equation*}
The basis of the lattice defined by the shifted carrier of the
equation \eqref{15.4} is composed of the vectors $(1,\,1)$,
$(5/4,\,0)$. The set $\mathbf{K}$ coincides with \eqref{13.11}.
Power expansions for $\eta^{(n,\,l,\,m,\,p)}(z)$ can be presented as
\begin{equation}
\begin{gathered}
\label{15.9}\eta^{(n,\,l,\,m,\,p)}(z)=q^{(n,\,l,\,m,\,p)}_{1/4}
z^{1/4}
+\sum_{k\,\in\,\,\mathbb{N}}q^{(n,\,l,\,m,\,p)}_{(1-5k)/4}\,z^{(1-5k)/4},\\
n,l=1,2,3,4;\,\, m=1,2,3;\,\, p=1,2.
\end{gathered}
\end{equation}
It can be shown that the coefficients $q^{(n,\,l,\,m,\,p)}_{-1}$
depend on $\alpha$. The exponential additions
$y^{(n,\,l,\,m,\,p)}(z)$ to the solutions $\xi^{(n,\,l,\,m)}(z)$ can
be written as
\begin{equation}
\begin{gathered}
\label{15.10}y^{(n,\,l,\,m,\,p)}(z)=C_3\,z^{q_{-1}^{(n,\,l,\,m,\,p)}}
\exp \left[\frac45\, q^{(n,\,l,\,m,\,p)}_{1/4}\,z^{5/4} +
\sum^{\infty}_{k=2} \frac{4}{5(1-k)} q^{(n,\,l,\,m,\,p)}_{(1-5k)/4}
z^{5(1-k)/4}\right],\\
n,l=1,2,3,4;\,\quad\, m=1,2,3;\,\quad\, p=1,2.
\end{gathered}
\end{equation}
They are exponentially small when
\begin{equation}
\label{15.11}Re
\left[q_{1/4}^{(n,\,l,\,m,\,p)}\,z^{5/4}\right]<0,\,\,\,z\rightarrow\infty.
\end{equation}

Thus for the solution expansions of the studied equation near the
point $z=\infty$ three-level exponential additions have been found.
Taking into account exponential additions, the solutions $w(z)$ at
$z\rightarrow\infty$ can be written as
\begin{equation*}
\begin{gathered}
\label{}w(z)=\varphi^{(n)}(z)+\exp\left[\int dz \left(
\zeta^{(n,\,l)}(z)+\exp \left[\int
dz\left(\xi^{(n,\,l,\,m)}(z)+\right. \right.\right.\right.\\
\left. \left. \left. \left. \exp\left(\int dz\,\eta^{(n,\,l,\,m,\,p)}(z)\right)\right)\right]\right)\right]
\end{gathered}
\end{equation*}
or as
\begin{equation}
\begin{gathered}
\label{15.12}w(z)=\varphi^{(n)}(z)+\exp\left[u^{(n,\,l)}(z)+\int dz
\,\exp \left[v^{(n,\,l,\,m)}(z)+\right.\right.\\
\left. \left. +\int dz\,\exp\left(y^{(n,\,l,\,m,\,p)}(z)\right)\right]\right],\\
n,l=1,2,3,4;\,\,\,m=1,2,3;\,\,\,p=1,2.
\end{gathered}
\end{equation}

Denote these solutions (in view of three-level additions) as
$G_2^{(1)}n\, G^{(1)}_1 l\,G^{(1)}_1 m\,G^{(1)}_1 p$ where
$n,l=1,2,3,4;\,\,\,m=1,2,3;\,\,\,p=1,2$.

\section{Exponential additions of the first level, corresponding to the edge $\Gamma^{(1)}_3$ at $\alpha\neq0$.}

Let us find the exponential addition of the first level $u(z)$ to
the expansion \eqref{6.5}, i.e. we will look for the solutions in
the form
\begin{equation}
\label{16.1} w(z)=\psi(z) + u(z).
\end{equation}
Everything written in section 13 up to the formula \eqref{13.7} is
true in the case of the edge $\Gamma^{(1)}_3$, taking into account
the only fact that instead of four expansions $\varphi^{(n)}(z)$ we
have only one $\psi(z)$. Thus we obtain the equation
\begin{equation}
\begin{gathered}
\label{16.2} \zeta_{zzz} + 4\zeta\zeta_{zz} +3(\zeta_z)^2 +6 \zeta^2
\zeta_z +  \zeta^4 -10\psi^2( \zeta_z + \zeta^2)-\\
-20\psi\psi_z\zeta-20\psi_z\psi_{zz}-10(\psi_z)^2+30 \psi^4 -z=0.
\end{gathered}
\end{equation}
Its carrier consists of the points
\begin{equation}
\begin{gathered}
\label{16.3}Q_{1,\,1} =(-3,1),\,\,\, Q_{1,\,2}=(-2,2),\,\,\, Q_{1,\,3}=(-1,3),\,\,\, Q_{1,\,4}=(0,4),\\
Q_{1,\,5}=(1,0),
\,\,\,Q_{2,\,k}=\left(-3-5k,1\right),\,\,\,Q_{3,\,k}=\left(-2-5k,2\right),\,\,\\
Q_{4,\,k}=\left(-4-5k,0\right),\,\,\,k \in \mathbb{N}\cup\{0\}.
\end{gathered}
\end{equation}
Convex closing of these points yields the half-string presented at
fig. 8.
\begin{figure}[h]
 \centerline{\epsfig{file=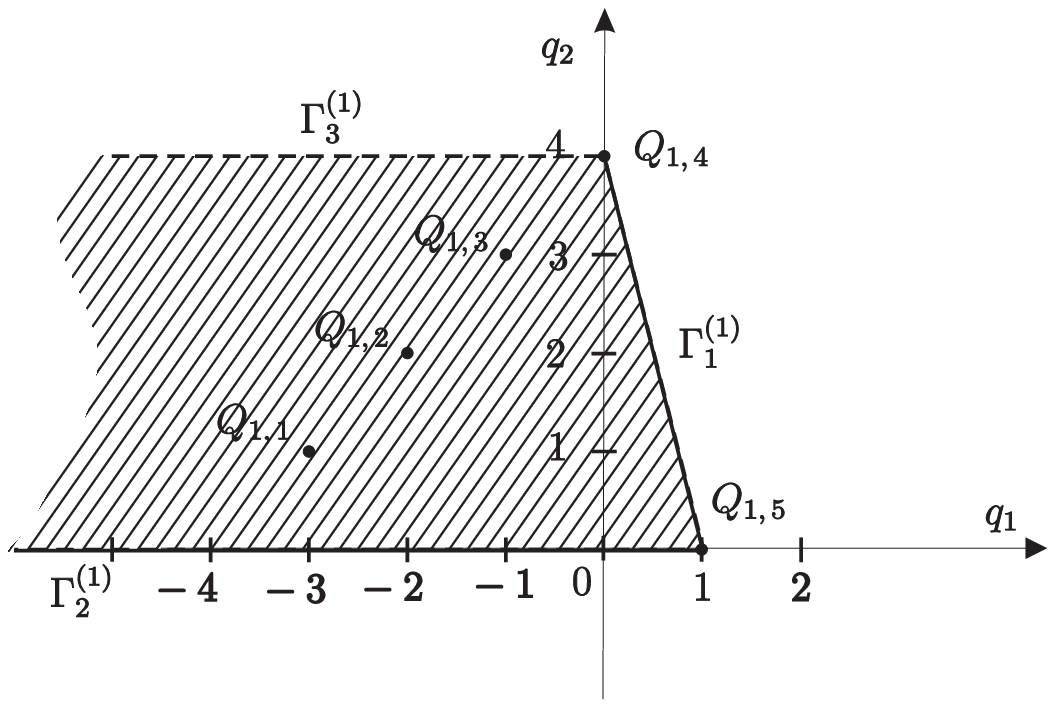,width=110mm}}
 \caption{}\label{fig:z_post}
\end{figure}
Its periphery is composed of two apexes $Q_{1,\,4}$, $Q_{1,\,5}$ and
of three edges $\Gamma^{(1)}_j\,\,(j=1,2,3)$ with the normal vectors
$N_1=(4,1),\,\, N_2=(0,-1),\,\, N_3=(0,1)$. The sufficient solutions
are given by the edge $\Gamma^{(1)}_1$. It is characterized by the
following reduced equation
\begin{equation}
\label{16.4}h_1^{(1)}(z,\zeta) \stackrel{def}{=} \zeta^4 -z=0.
\end{equation}
Equation \eqref{16.4} has four solutions
\begin{equation}\begin{gathered}
\label{16.5}\zeta^{(l)}(z)
=g_{1/4}^{(l)}z^{1/4},\,\,\,l=1,2,3,4;\,\,\,g_{1/4}^{(l=\{1,\,2,\,3,\,4\})}=1,-1,i,-i.
\end{gathered}
\end{equation}
The shifted carrier of the equation \eqref{16.2} lies in the lattice
with the basis $(-1,4),\,(-2,3)$. Together with the shifted carrier
of the reduced solutions \eqref{16.5} they generate the new lattice
with the basis $(-1/4,1),\,(-2,3)$. The cone of the problem is
$\mathcal{K}=\left\{k<\frac14\right\}$, then we get
\begin{equation}
\label{16.6}\mathbf{K}=\left\{\frac{1-5k}4,k\in\mathbb{N}\right\}.
\end{equation}
Power expansions for the reduced solutions \eqref{16.5} can be
written as
\begin{equation}\begin{gathered}
\label{16.7}\zeta^{(l)}(z)=g^{(l)}_{1/4} z^{1/4}
+\sum_{k\,\in\,\,\mathbb{N}}g^{(l)}_{(1-5k)/4}\,z^{(1-5k)/4},\,\,\,
 l=1,2,3,4.
\end{gathered}\end{equation}
In this expression $g^{(l)}_{-1}=-3/8$. Using the formula
\eqref{13.13} we get exponential additions
\begin{equation}
\begin{gathered}
\label{16.8}u^{(l)}(z)=\frac{\tilde{C}_1}{z^{3/8}}\, \exp
\left[\frac45\, g^{(l)}_{1/4}\,z^{5/4} + \sum^{\infty}_{k=2}
\frac{4}{5(1-k)} g^{(l)}_{(1-5k)/4} z^{5(1-k)/4}\right], \,\,
l=1,2,3,4.
\end{gathered}
\end{equation}
Denote them as $G_3^{(1)}\, G^{(1)}_1 l$. Here and later
$\tilde{C}_1,\,\tilde{C}_2,\,\tilde{C}_3$ are the arbitrary
constants. Taking into consideration first two members of the series
in \eqref{16.8} we can rewrite this expression in the form
\begin{equation*}
\begin{gathered}
\label{}u^{(l)}(z)=\frac{\tilde{C}_1}{z^{3/8}}\, \exp
\left[\frac45\,g^{(l)}_{1/4}{z}^{5/4}-\,\frac
{64\,\alpha^2-9}{32g^{(l)}_{1/4}{z}^{5 /4}}-\,\frac
{5(64\,\alpha^2-9)}{256(g^{(l)}_{1/4})^2 z^{5/2}}+...\right], \,\,
l=1,2,3,4.
\end{gathered}
\end{equation*}
The applicability condition \eqref{13.15} holds for the additions
\eqref{16.8} provided that $g^{(n,\,l)}_{1/4}$ replace
$g^{(l)}_{1/4}$.

\section{Exponential additions of the second level, corresponding to the edge $\Gamma^{(1)}_3$ at $\alpha\neq0$.}

Now we will look for exponential additions of the second level
$v^{(l)}(z)$, i.e. we will look for the additions to the solutions
$\zeta^{(l)}(z)$. All transformations of the section 14, which
converted the equation \eqref{14.1} to the equation \eqref{14.4} do
not change under the condition $v\stackrel{def}{=} v^{(l)}$ è
$\zeta\stackrel{def}{=} \zeta^{(l)}$. As a result we get the
equation
\begin{equation}
\begin{gathered}
\label{17.1}\xi_{zz} + 3\, \xi\,\xi_{z} + \xi^3
+4\zeta(\xi_z+\xi^2)+6(\zeta_z+\zeta^2)\xi -10\,\psi^2\xi+
4\zeta_{zz} + 12 \zeta\zeta_z  +
\\+4\,\zeta^3  -20\, \psi^2\,\zeta
-20\,\psi\,\psi_z=0.
\end{gathered}
\end{equation}
The carrier of this equation differs from the carrier of the
equation \eqref{14.4}:
\begin{equation}
\begin{gathered}
\label{17.2}M_{1,\,1}=(-2,1),\,\,\, M_{1,\,2}=(-1,2),\,\,\,
M_{1,\,3}=(0,3),\,\,\,
M_{2,\,k_1}=\left(-\frac34-\frac54k_1,\,1\right),\,\,\,\\
M_{3,\,k_1}=\left(\frac14-\frac54k_1,2\right),\,\,\,M_{4,\,k_1}=\left(\frac12-\frac54k_1,1\right),\,\,\,M_{5,\,k_1}=\left(-2-5k_1,\,1\right),\,\,\,\\
M_{6,\,k_1}=\left(-\frac74-\frac54k_1,\,0\right),\,\,\,M_{7,\,k_1}=\left(-\frac12-\frac54k_1,\,0\right),\,\,
\,M_{8,\,k_1}=\left(\frac34-\frac54k_1,\,0\right),\,\, \,\\
M_{9,\,k_1}=\left(-3-5k_1,\,0\right),\,\, \,
M_{10,\,k_1,\,k_{2}}=\left(-\frac74-5k_1-\frac54k_{2},\,0\right),\,\,
\,k_{i} \in \mathbb{N}\cup\{0\},\, i=1,2.
\end{gathered}
\end{equation}
The convex hull obtained after locking of these points is similar to
the one described in the section 14. The reduced equation
corresponding to the edge $\Gamma_1^{(1)}$ is the following
\begin{equation}
\begin{gathered}
\label{17.3}\xi^3 + 4g^{(l)}_{1/4}z^{1/4}\xi^2 + 6(g^{(l)}_{1/4})^2
z^{1/2}\xi + 4(g^{(l)}_{1/4})^3 z^{3/4} =0.
\end{gathered}
\end{equation}
The equation \eqref{17.3} has $12$ solutions
\begin{equation}
\begin{gathered}
\label{17.4}\xi^{(l,\,m)}(z)=r_{1/4}^{(l,\,m)}\, z^{1/4},\,\,\,\,
l=1,2,3,4;\,\,\,\, m=1,2,3,
\end{gathered}
\end{equation}
where calculation of the coefficients $r_{1/4}^{(l,\,m)}$ yields
\begin{equation*}
\begin{gathered}
\label{}r_{1/4}^{(l,\,1)}\,=-2g_{1/4}^{(l)};\,
r_{1/4}^{(l,\,2)}\,=(-1+i)\,g_{1/4}^{(l)};\,r_{1/4}^{(l,\,3)}\,=-(1+i)\,g_{1/4}^{(l)},\,\,\,\,
l=1,2,3,4.
\end{gathered}
\end{equation*}
The lattice which contains the shifted carriers of the equation
\eqref{17.1} and the reduced solutions \eqref{17.4} coincides with
the lattice of the section 13. That is why power expansions for the
reduced solutions \eqref{17.4} are
\begin{equation}
\begin{gathered}
\label{17.5} \xi^{(l,\,m)}=r_{1/4}^{(l,\,m)} z^{1/4}
+\sum_{k\,\in\,\,\mathbb{N}}r^{(l,\,m)}_{(1-5k)/4}\,z^{(1-5k)/4},\quad\,l=1,2,3,4;\,\quad\,
m=1,2,3.
\end{gathered}
\end{equation}
Here $r_{-1}^{(l,\,m)}=1/4$. Returning to the variables
$v^{(l,\,m)}(z)$ we get the exponential additions $G_3^{(1)}\,
G^{(1)}_1 l\,G^{(1)}_1 m$ and their applicability condition
\begin{equation}
\begin{gathered}
\label{17.6}v^{(l,\,m)}(z)=\tilde{C}_2\,z^{1/4}\, \exp
\left[\frac45\, r_{1/4}^{(l,\,m)}\,z^{5/4} + \sum^{\infty}_{k=2}
\frac{4}{5(1-k)}
r^{(l,\,m)}_{(1-5k)/4}z^{5(1-k)/4}\right];\\
Re\left[r_{1/4}^{(l,\,m)}\,z^{5/4}\right]<0,\,\,z\rightarrow\infty;
\,\quad\,l=1,2,3,4;\,\quad\,m=1,2,3.
\end{gathered}
\end{equation}

\section{Exponential additions of the third level, corresponding to the edge $\Gamma^{(1)}_3$ at $\alpha\neq0$.}

Exponential additions of the third level $y^{(l,\,m)}(z)$ (additions
to the solutions $\xi^{(l,\,m)}(z)$) can be found as they have been
found for the edge $\Gamma^{(1)}_2$ in the section 15. Let us in
brief follow the procedure of calculations. After essential
transformations we get the equation
\begin{equation}
\begin{gathered}
\label{18.1}\eta_{z} +\eta^2+ (3\xi +
4\zeta)\eta+3\xi_z+3\xi^2+8\xi\zeta+6\zeta_z+6\zeta^2-10\psi^2=0.
\end{gathered}
\end{equation}
Its carrier consists of the following points
\begin{equation}
\begin{gathered}
\label{18.2}R_{1,\,1}=(-1,\,1),\,\,\, R_{1,\,2}=(0,\,2),\,\,\, R_{2,\,k}=(-\frac34-\frac54k,\,0),\,\,\, R_{3,\,k}=\left(\frac12-\frac54k,\,0\right),\\
R_{4,\,k}=\left(-2-5k,\,0\right),\,\, \,
R_{5,\,k}=\left(\frac14-\frac54k,\,1\right),\,\, \,k \in
\mathbb{N}\cup\{0\}.
\end{gathered}
\end{equation}
Closing of their convex hull yields the half-string similar to the
one of the section 15. The edge $\Gamma_1^{(1)}$ is characterized by
the reduced equation
\begin{equation}
\begin{gathered}
\label{18.3}\eta^2+ (3r_{1/4}^{(l,\,m)}+4g_{1/4}^{(l)})z^{1/4}\eta+
\left(3(r_{1/4}^{(l,\,m)})^2+8r_{1/4}^{(l,\,m)}g_{1/4}^{(l)}+
6(g_{1/4}^{(l)})^2 \right)z^{1/2}=0
\end{gathered}
\end{equation}
which possesses $24$ solutions
\begin{equation}
\begin{gathered}
\label{18.4}\eta^{(l,\,m,\,p)}(z)=q^{(l,\,m,\,p)}_{1/4}\,
z^{1/4};\,\,\, l=1,2,3,4;\,\,\,\, m=1,2,3; \,\,\,\,p=1,2.
\end{gathered}
\end{equation}

The coefficients $q^{(l,\,m,\,p)}_{1/4}$ are equal to
\begin{equation*}
\begin{gathered}
\label{}
q_{1/4}^{(l,\,m,\,p=\{1,\,2\})}=-\frac32\,r_{1/4}^{(l,\,m)}-2\,g_{1/4}^{(l)}
\pm\,\frac12i\sqrt{{(3\,(r_{1/4}^{(l,\,m)})^{2}+8\,{r_{1/4}^{(l,\,m)}}\,g_{1/4}^{(l)}+8\,
(g_{1/4}^{(l)})^{2}}},\\
 l=1,2,3,4;\,\,\,\,m=1,2,3.
\end{gathered}
\end{equation*}
The set $\mathbf{K}$ in this case is \eqref{13.11}. Consequently
power expansions for the solutions \eqref{18.4} can be written as
\begin{equation}
\begin{gathered}
\label{18.5}\eta^{(l,\,m,\,p)}(z)=q^{(l,\,m,\,p)}_{1/4} z^{1/4}
+\sum_{k\,\in\,\,\mathbb{N}}q^{(l,\,m,\,p)}_{(1-5k)/4}\,z^{(1-5k)/4},\,\,\,\\
l=1,2,3,4;\,\quad\, m=1,2,3;\,\quad\, p=1,2.
\end{gathered}
\end{equation}
The coefficients $q^{(l,\,m,\,p)}_{-1}$ do not depend on $\alpha$
and $q^{(l,\,m,\,p)}_{-1}=1/4$.

So the exponential additions $G_3^{(1)}\, G^{(1)}_1 l\,G^{(1)}_1
m\,G^{(1)}_1 p$ are the following
\begin{equation}
\begin{gathered}
\label{18.6}y^{(l,\,m,\,p)}(z)=\tilde{C}_3\,z^{1/4} \exp
\left[\frac45\, q^{(l,\,m,\,p)}_{1/4}\,z^{5/4} + \sum^{\infty}_{k=2}
\frac{4}{5(1-k)} q^{(l,\,m,\,p)}_{(1-5k)/4}
z^{5(1-k)/4}\right],\\
l=1,2,3,4;\,\quad\, m=1,2,3;\,\quad\, p=1,2.
\end{gathered}
\end{equation}

The applicability condition for the additions \eqref{18.6} coincides
with \eqref{15.11} provided that $q_{1/4}^{(n,\,l,\,m,\,p)}$ replace
$q_{1/4}^{(l,\,m,\,p)}$. Taking into account obtained additions the
solution $w(z)$ can be written as it has been done in formula
\eqref{15.12}.

\section{Conclusion.}

In this work the fourth-order analogue to the second Painlev\'{e}
equation was studied with a help of the power geometry method
\cite{Bruno01, Bruno02, Bruno03}. We found all power and non-power
asymptotics of its solutions, power expansions generated by these
power asymptotics and exponential additions which correspond to
certain expansions. The results depend on the value of parameter
$\alpha$.

First of all let us briefly review the case $\alpha \neq0$. Near the
point $z=0$ we obtained one-parametric family $G_1^{(0)}4$, three
two-parametric families $G_1^{(0)}3$, $G_1^{(1)}3$, $G_1^{(1)}4$,
three three-parametric families $G_1^{(0)}2$, $G_1^{(1)}1$,
$G_1^{(1)}2$, one four-parametric family $G_1^{(0)}1$ and the family
$G_4^{(1)}1$ (the families $G_1^{(0)}2$, $G_1^{(0)}3$, $G_1^{(0)}4$,
$G_4^{(1)}1$ are the special cases of $G_1^{(0)}1$). These
expansions converge for small $|z|$. Their existence and analyticity
follow from the Cauchy theorem.

Besides near the point $z=\infty$ we found the families:
$G_2^{(1)}n\,\,(n=1,2,3,4)$ and $G_3^{(1)}1$. For each of these
expansions three-level exponential additions we computed:
$G_2^{(1)}n\, G^{(1)}_1 l\,G^{(1)}_1 m\,G^{(1)}_1
p\,\,\,(n,l=1,2,3,4;\,\,\,m=1,2,3;\,\,\,p=1,2)$ and $G_3^{(1)}\,
G^{(1)}_1 l\,G^{(1)}_1 m\,G^{(1)}_1 p\,\,\,
(l=1,2,3,4;\,\,\,m=1,2,3;\,\,\,p=1,2)$.

In addition it is important to mention that some partial solutions
of the studied equation can be found using the expansion
$G_3^{(1)}1$. It is obvious that at the certain values of the
parameter $\alpha$ (more exactly $\alpha=-2,\,\,-1,\,\,1,\,\,2$) the
series in the formula \eqref{6.5} truncates and we get rational
solutions
\begin{equation}
\label{19.1}w(z)=\frac{-\alpha}{z},\,\,\,\,  \alpha=-2,-1,\,1,\,2.
\end{equation}
Application of these solutions as "seed solutions" in the
B\"{a}cklund transformations for the studied equation yields other
rational solutions at whole values of the parameter $\alpha$.
Partial solutions \eqref{19.1} can be also obtained from the
expansions $G_1^{(1)}1$, $G_1^{(1)}2$, $G_1^{(1)}3$ and
$G_1^{(1)}4$.

In the case $\alpha=0$ near the point $z=0$ we found one-parametric
family $\tilde{G}_1^{(0)}4$, three two-parametric families
$\tilde{G}_1^{(0)}3$, $\tilde{G}_1^{(1)}3$, $\tilde{G}_1^{(1)}4$,
three three-parametrical families $\tilde{G}_1^{(0)}2$,
$\tilde{G}_1^{(1)}1$, $\tilde{G}_1^{(1)}2$ and one four-parametric
family $\tilde{G}_1^{(0)}1$ of power expansions (the families
$\tilde{G}_1^{(0)}2$, $\tilde{G}_1^{(0)}3$, $\tilde{G}_1^{(0)}4$ are
the special cases of $G_1^{(0)}1$). All of them converge for small
$|z|$.

Near the point $z=\infty$ we computed four families of non-power
asymptotics $\tilde{G}_3^{(1)}n,\,\,(n=1,2,3,4)$ and four families
of power expansions $\tilde{G}_2^{(1)}n\,\,(n=1,2,3,4)$.

The expansions $G_1^{(1)}n\,\,(n=1,2,3,4)$ were found earlier
\cite{Kudryashov01,Kudryashov10}, while all other expansions are the
new ones.

To sum up we would like to emphasize that the obtained power
expansions differ from the power expansions of the Painlev\'{e}
equations $P_1\div P_6$ solutions  \cite{Bruno04, Bruno05, Bruno06,
Bruno07, Gromak01},. This fact can be interpreted as the additional
proof of the hypothesis that the fourth-order equation \eqref{1.3}
determines new transcendental functions as the equations $P_1\div
P_6$ do.


\begin{thebibliography}{99}


\bibitem{Ablowitz01} \textit{Ablowitz M.J., Clarcson P.A.} Solitons, Nonlinear Evolution Equations and Inverse Scattering.
Cambridge University Press; 1991.

\bibitem{Barouch03} \textit{Barouch E., McCay B.M., Wu T.T.} Phys Rev Lett 1973; 31.

\bibitem{Brezin04} \textit{Brezin E., Kazakov V.} Phys Lett B 1990;236:144-150.

\bibitem{De_Boer05} \textit{De Boer P.C.T., Ludford L.S.S.} Plazm Phys 1975;17:29-43.

\bibitem{Ablowitz06} \textit{Ablowitz M.J., Segur H.} Phys Rev Lett 1997;38:1103-1106.

\bibitem{Hall07} \textit{Hall P.} IMA J Appl Math 1982;29:173-196.

\bibitem{Chandrasecar08} \textit{Chandrasekar S.} Proc Roy Soc London A 1986;408:209-232.

\bibitem{Kudryashov01} \textit{Kudryashov N.A.} Analytical theory of nonlinear differential
equations,  Institute of Computer Investigations, Moscow-Igevsk,
2004, 360 p. (in Russian).

\bibitem{Kudryashov02} \textit{Kudryashov N.A.} Phys
Lett A 1997;233:387-400.

\bibitem{Kudryashov03} \textit{Kudryashov N.A.} Phys Lett A 1997;224 N 6.
P. 353--360.

\bibitem{Kudryashov04} \textit{Kudryashov N.A.} J Phys A: Math Gen 1998;31:N 6.
P. L.129--L.137.

\bibitem{Kudryashov05} \textit{Kudryashov N.A., Soukharev M.B.} Phys Lett A 1998;237:206-216.

\bibitem{Kudryashov06} \textit{Kudryashov N.A., Pickering A.} J Phys A: Math Gen 1998;31:999 - 1014.

\bibitem{Kudryashov07} \textit{Kudryashov N.A.} Phys Lett A: 1999;252:173-179.

\bibitem{Kudryashov08} \textit{Kudryashov N.A.} J Phys A: Math Gen 1999;32:999--1013.

\bibitem{Kudryashov09} \textit{Kudryashov N.A.} Phys Lett A 2000;273:194 -- 202,353--360.

\bibitem{Kudryashov10} \textit{Kudryashov N.A.} Theoretical and mathematical physics 2000;122:72 - 86.

\bibitem{Kudryashov11} \textit{Kudryashov N.A., Pickering A.} CRM Proceedings and Lecture Notes 2000;25:245-253.

\bibitem{Kudryashov12} \textit{Kudryashov N.A.} J Phys A: Math Gen 2002;35:93 -- 99.

\bibitem{Kudryashov13} \textit{Kudryashov N.A.} J Phys A: Math Gen 2002;35:4617--4632.

\bibitem{Kudryashov14} \textit{Kudryashov N.A., Soukharev M.B.} ANZIAM Industrial and Applied Mathematics 2002.

\bibitem{Kudryashov15} \textit{Kudryashov N.A.} J Math Phys 2003;44:6160--6178.

\bibitem{Kudryashov16} \textit{Kudryashov N.A., Efimova O.Yu.} Chaos, Solitons \& Fractals; 2006 (in press).

\bibitem{Airault01} \textit{Airault H.} Studies in applied Mathematics 1979;61:31 -- 53.

\bibitem{Clarkson01} \textit{Clarkson P.A., Joschi N., Pickering A.} Inverse problems 1999;15:175 -- 187.

\bibitem{Clarkson02} \textit{Clarkson P.A., Hone A.N.W., Joschi N.} Journal of Nonlinear Mathematical
physics 2003;10.

\bibitem{Cosgrove01} \textit{Cosgrove C.M.} Study Appl Math 2000;104:1 -- 65.

\bibitem{Joshi01} \textit{Creswell G., Joshi N.} J Phys A: Math Gen 1999;32:655 - 669.

\bibitem{Hone01} \textit{Hone Andrew N.W.} Physica D 1998;118:1 -- 16.

\bibitem{Hone02} \textit{Hone Andrew N.W.} J Phys A 2001;34:2235 -- 2245.

\bibitem{Gordoa01} \textit{Gordoa P.R., Pickering A.} Journal of mathematical physics 1999;11:5749 -- 5766.

\bibitem{Gordoa02} \textit{Gordoa P.R.}Phys Lett A 2001;287:365 - 370.

\bibitem{Flashka01} \textit{Flaschka H., Newell A.C.} Communications in Mathematical Physics 1980;76:65 -- 116.

\bibitem{Kawai01} \textit{Kawai T., Koike T., Nishikawa Y., Takei Y.} 2004 preprint/RS/RIMS 1471,Kioto.

\bibitem{Mugan01} \textit{Mugan U., Jrad F.} J Phys A: Math Gen 1999;32:7933 - 7952.

\bibitem{Mugan02} \textit{Mugan U., Jrad F.} Journal of Nonlinear Mathematical Physics A 2002;9(3):282-310.

\bibitem{Mugan03} \textit{Mugan U., Jrad F.} Zaitschrift fur Naturforshing 2004;9 (3):282-310.

\bibitem{Mugan04} \textit{Mugan U., Jrad F.} Zaitschrift fur Naturforshung 2005;9(3):282-310.

\bibitem{Nijhoff01} \textit{Nijhoff F.W., Walker A.J.} Glasgow Math J 2001;43A:199 -- 123.

\bibitem{Pickering01} \textit{Pickering A.} Phys Lett A 2002;301:275 - 280.

\bibitem{Shimomura01} \textit{Shimomura S.} Proc Japan Acad 80 Ser A 2004:105 -- 109.

\bibitem{Bruno01} \textit{Bruno A.D.} Power geometry in algebraic and differential equations,
Moscow, Nauka, Fizmatlit, 1998, 288 p (in Russian).

\bibitem{Bruno02} \textit{Bruno A.D. } ISAAC; 2001;51-71.

\bibitem{Bruno03} \textit{Bruno A.D.} Uspehi of mathematical nauk 2004;59:31-80.

\bibitem{Bruno04} \textit{Bruno A.D., Petrovich V.Yu.} KIAM preprint 2004;No.9,Moscow(in Russian).

\bibitem{Bruno05} \textit{Bruno A.D., Zavgorodnya Yu.B.} KIAM preprint 2003;No.48,Moscow(in Russian).

\bibitem{Bruno06} \textit{Bruno A.D., Karulina E.S.} Doklady RAN 2004;395,No.4:439 -- 444(in Russian).

\bibitem{Bruno07} \textit{Bruno A.D., Goruchkina I.B.} Doklady RAN 2004;395,No.6:733-737(in Russian).

\bibitem{Gromak01} \textit{Gromak V.I., Laine I., Shimomura S.} Painleve Differential Equations in
the Complex Plane, Walter de Gruyter, Berlin, New York, 2002.

\end{thebibliography}
\end{document}